\documentclass[12pt]{nature}
\usepackage{graphicx}
\usepackage{amssymb}
\usepackage{amsmath}
\usepackage{mathrsfs}
\usepackage{bm}
\usepackage{color}
\usepackage{rotating} 
\usepackage{empheq}
\usepackage{cases}
\usepackage{subeqnarray}

\usepackage{amsfonts}
\usepackage{mathrsfs}
\usepackage{bm}
\usepackage{lipsum}
\usepackage{physics}
\usepackage{multirow}
\usepackage{rotating}
\usepackage{graphicx}
\usepackage{soul}
\usepackage{longtable,lscape,threeparttablex,booktabs}

\newcommand{\aap}{{\it Astron. Astrophys.}}
\newcommand{\araa}{{\it Annu. Rev. Astron. Astrophys.}}
\newcommand{\apj}{{\it Astrophys. J.}}
\newcommand{\aj}{{\it Astron. J.}}
\newcommand{\apjl}{{\it Astrophys. J. Letters}}
\newcommand{\apjs}{{\it Astrophys. J. Supp.}}
\newcommand{\nat}{{\it Nature}}

\newcommand{\pasp}{{\it Pub. Astron. Soc. Pac.}}
\newcommand{\pasa}{{\it Pub. Astron. Soc. Australia}}

\newcommand{\prd}{{\it Phys. Rev. D}}
\newcommand{\mnras}{{\it Mon. Not. R. Astron. Soc.}}

\newcommand{\aapr}{{\it Astron. Astrophys. Rev.}}
\newcommand{\physrep}{{\it Phys. Rep.}}
\newcommand{\prl}{{\it Phys. Rev. Lett}}

\def\bmy{\bm{y}}
\def\bmnn{\bm{n}}
\def\bmC{\bm{C}}
\def\bmE{\bm{E}}
\def\bmS{\bm{S}}
\def\bmN{\bm{N}}
\def\bms{\bm{s}}
\def\bhm{M_{\bullet}}

\def\bmn{\bm{n}_{\rm obs}}

\def\calD{{\mathscr{D}}}
\def\calO{{\cal O}}
\def\calR{{\cal{R}}_{0}}

\def\ergs{\rm ergs\,s^{-1}}
\def\fline{\ell_{\!\lambda}}
\def\feii{Fe\,{\sc ii}}
\def\Fline{F_{\ell}}
\def\kms{\rm km\,s^{-1}}
\def\nobs{\bm{n}_{\rm obs}}

\def\RBLR{R_{\rm BLR}}

\def\sigmad{\sigma_{\rm d}}
\def\sunm{M_{\odot}}
\def\taud{\tau_{\rm d}}
\def\xiBLR{\xi_{\rm BLR}}

\begin{document}

\baselineskip 20.5pt




\title{A parallax distance to 3C 273 through spectroastrometry and reverberation mapping}

\author{
Jian-Min Wang$^{1,2,3}$, 
Yu-Yang Songsheng$^{1,2}$,
Yan-Rong Li$^1$,
Pu Du$^1$ and
Zhi-Xiang Zhang$^{4}$
}

\date{}

\maketitle

\begin{affiliations}
\item{Key Laboratory for Particle Astrophysics, Institute of High Energy Physics,
Chinese Academy of Sciences, 19B Yuquan Road, Beijing 100049, China}

\item{School of Astronomy and Space Science, University of Chinese Academy of Sciences, 
19A Yuquan Road, Beijing 100049, China}  

\item{National Astronomical Observatories of China, Chinese Academy of Sciences,
 20A Datun Road, Beijing 100020, China}
 
\item{Department of Astronomy, Xiamen University, Xiamen, Fujian 361005, China} 
\end{affiliations}

\begin{abstract}
Distance measurements for extragalactic objects are a fundamental problem in 
astronomy\cite{Peacock1999,deGrijs2011} and cosmology\cite{Freedman2010,Weinberg2013}.
In the era of precision cosmology, we urgently need better measurements of cosmological 
distances to observationally test the increasing $H_{0}$ tension of the Hubble constant 
measured from different tools\cite{Freedman2017,Aghanim2018,Riess2019}.
Using spectroastrometry\cite{Abuter2017}, GRAVITY at The Very Large Telescope 
Interferometer successfully revealed the structure, kinematics and angular sizes
of the broad-line region (BLR) of 3C 273 with an unprecedentedly high spatial 
resolution\cite{Sturm2018}. Fortunately, reverberation mapping (RM)\cite{Blandford1982} 
of active galactic nuclei (AGNs) reliably provides linear sizes of their 
BLRs\cite{Peterson1993}. Here we report a joint analysis of spectroastrometry and RM 
observations to measure AGN distances. We apply this analysis to 
3C 273 observed by both GRAVITY\cite{Sturm2018} and an RM campaign\cite{Zhang2019}, 
and find an angular distance of $551.5_{-78.7}^{+97.3}$\,Mpc and 
$H_{0}=71.5_{-10.6}^{+11.9}\,{\rm km\,s^{-1}\,Mpc^{-1}}$. Advantages of 
the analysis are 1) its pure geometrical measurements and 2) 
it simultaneously yields mass of the central 
black hole in the BLR. Moreover, we can conveniently repeat measurements 
of selected AGNs to efficiently reduce the statistical and systematic errors. 
Future observations of a reasonably sized sample ($\sim 30$ AGNs) will 
provide distances of the AGNs and hence a new way of 
measuring $H_{0}$ with a high precision $\left(\lesssim 3\%\right)$ 
to test the $H_{0}$ tension.
\end{abstract}

GRAVITY observations through spectroastrometry (SA) sensitively detect the angular 
structure of the BLR in a 
direction perpendicular to line-of-sight (LOS) whereas RM observations are more sensitive 
along the direction of the sight. A joint analysis of SA and RM observations 
of AGNs (hereafter SARM) can thus directly measure absolute angular distances ($D_{\rm A}$).
AGN emission lines arise from the photoionization of clouds by the central energy 
source\cite{Osterbrock1986,Osterbrock1989,Ho2008}. An assembly of ionized clouds orbiting 
around the central black hole with mass ($\bhm$), in which emission lines are broadened 
by the Doppler motion in the black hole gravity, is the well-known paradigm of the BLR. 
The SA measures the wavelength dependence of the photocentre of an 
object so that it provides information on the spatial structure of the object
on scales much smaller than the diffraction limit\cite{Bailey1998}. For an interferometer 
with a baseline $\bm{B}$, a non-resolved source with a global angular size smaller than 
its resolution limit $\lambda / B$ has the interferometric phase
\begin{equation}\label{eq:phase}
\phi_*(\lambda,\lambda_{\rm r})=-2\pi\bm{u}\vdot[\bm{\epsilon}(\lambda)-\bm{\epsilon}(\lambda_{\rm r})],
\end{equation}
where $\bm{u}=\bm{B}/\lambda$ is the spatial frequency,
$\bm{\epsilon}$ is the photocentre of the source at wavelength $\lambda$ and $\lambda_{\rm r}$ 
is the wavelength of a reference channel. Here the bold letters are vectors. 
We use an established approach\cite{Petrov1992,Rakshit2015} to calculate differential phase 
curves (see also the Methods for details) for the angular sizes of the BLR  
if the BLR geometry is given for general cases. 3C 273, the first 
quasar discovered\cite{Schmidt1963}, has a redshift of $z=0.158$ and a $K$-band magnitude 
$K\approx 10.0$ (both from the NASA Extragalactic Database).
GRAVITY successfully measured the differential phase curves ($\phi_{*}$) of 3C 273 
in July 2017, January, March, and May 2018\cite{Sturm2018}. The spectral resolution of 
GRAVITY is $\lambda/\Delta\lambda\approx 500$, which is good enough to constrain
some parameters of the BLR from the observed Paschen $\alpha$ (hereafter Pa$\alpha$) line. 
The present analysis is based on GRAVITY data.

As a consequence of photoionization, the broad emission lines
respond to variations of the ionizing continuum with delays (usually the optical 
5100\AA\, continuum is used as a proxy). It has been observationally demonstrated that 
the RM technique measures the time lags of the lines with respect to the
continuum and provides absolute sizes of the BLRs\cite{Peterson1993}. 
3C 273 is known as a blazar with a powerful jet\cite{Courvoisier1998}, but it has a 
prominent big blue bump dominating from optical to soft X-rays over the non-thermal 
emissions of the jet\cite{Walter1994}. The varying continuum is
contaminated sometimes to some degrees by the jet's emission, but the prominent 
big blue bump emission dominates to govern H$\beta$ reverberation most of the 
time. A 10-yr RM campaign of 3C 273 has been conducted through joint observations
on the Bok 2.3m telescope at Steward Observatory, University of Arizona, and the Lijiang 
2.4m telescope in Yunnan Observatory, Chinese Academy of Sciences\cite{Zhang2019}. 
The campaign using the Bok telescope started from March 2008 and the Lijiang telescope 
from December 2016 to May 2018. We have 296 spectra with a mean cadence of 
7.4\,days for the entire campaign. One comparison star was simultaneously observed 
in a long slit with 3C 273 and used for flux calibrations. This method 
generates high-quality light curves (LCs), with the H$\beta$ flux having a typical error at 
a $\sim 2\%$ level and the continuum at $\sim 1\%$. Details of the campaign and data reduction 
can be found in Ref.\cite{Zhang2019}, from which this joint analysis takes the data.

Reverberation of broad emission lines delivers the
linear sizes of the emitting regions while spectroastrometry probes their angular sizes,
however, their geometrical sizes remain open without a reliable physical model of the 
regions,
as they are usually explained as a kind of emissivity-averaged sizes or mean centers of 
$\lambda$-wavelength photons, respectively. We have to specify a BLR model for 
the joint analysis when we combine GRAVITY and RM data.
Many efforts have been made to model the RM data of $\sim 40$ mapped AGNs in details
through Markov Chain Monte Carlo simulations\cite{Pancoast2011,Li2013}, offering 
empirical formulations of spatial distributions of the BLR clouds for the present joint 
analysis. We follow the approach described in Ref.\cite{Pancoast2011,Li2013,Pancoast2014} 
for RM$^{\rm 1D}$ modeling (see details in the Methods), but we take the simplest version 
of the current model by keeping necessary parameters, which are listed in Table 1. 
Here the roles of individual parameters in GRAVITY and RM data 
are also highlighted for the necessity of the joint analysis of GRAVITY and RM data.
It is important to note that the joint analysis can simultaneously generate 
the distances and the central black hole masses of AGNs.

Modeling of RM data shows that the radial structure is described by a shifted 
$\Gamma$-distribution\cite{Pancoast2014}. The distance of BLR clouds from the SMBH
is computed by
$r = R_{\rm S} + \mathscr{F}R_{\rm BLR} +\Gamma_{0} \beta^{2} (1-\mathscr{F})R_{\rm BLR}$,
where $R_{\rm S}=2G\bhm/c^{2}$ is the Schwarzschild radius,
$G$ is the gravitational constant, $c$ is the speed of light, $R_{\rm BLR}$ is the mean radius,
$\mathscr{F}=R_{\rm in}/R_{\rm BLR}$ is the fraction of the inner to the mean radius,
and $\beta$ is the shape parameter. Here $\Gamma_{0} = p(x|\beta^{-2},1)$ is a random number drawn 
from a $\Gamma$-distribution 
$	p(x | \alpha, x_{0})=x^{\alpha-1} \exp (-x / x_{0})/x_{0}^{\alpha} \Gamma(\alpha)$,
where $x_{0}$ is a scale factor, $\alpha=\beta^{-2}$, and $\Gamma(\alpha)$ is the $\Gamma$-function. 
Such a radial distribution of BLR clouds is convenient for calculations and naturally covers
several simple cases\cite{Pancoast2014}. On the other hand, multiple campaigns of several AGNs show 
Keplerian rotation of the BLR clouds around the central black hole (see the Methods for details of the
references). Moreover, this is also directly supported by the differential phase curves\cite{Sturm2018} 
of 3C 273 which is also in agreement with the velocity-resolved delays\cite{Zhang2019}. In this paper, 
we employ a Keplerian disc with an opening angle as the BLR model in 3C 273 for this joint analysis
(see Supplementary Figure 1). 
 
The joint analysis employs three datasets, which are 1) the long term RM data; 2) the differential 
phase curves; 3) the Pa$\alpha$ line profiles. The analysis can be conducted by maximizing the 
posterior probability distributions of the model parameters for the SARM data. We assume that the 
probability distributions for the measurement values of LCs, profiles and differential phase curves 
are Gaussian and uncorrelated.  Accordingly, we generate their corresponding probabilities for fitting 
observational points of each dataset, which are $P^{f_{\ell}}_{i}$ for flux variations of H$\beta$ line,  
$P_{i,j}^{\phi}$ for differential phase curves and $P_{j}^{F_{\ell}}$ for profiles of the Pa$\alpha$ 
line, respectively (given in the Methods). The joint likelihood function can be expressed by productions 
of the three probabilities,
\begin{equation}\label{eq:probability}
P(\calD|\Theta)=\prod_{i=1}^{N_{\rm RM}}P^{f_{\ell}}_{i}\times \prod_{i=1}^{N_{\rm G}} \prod_{j=1}^{N_{\lambda}}
                P_{i,j}^{\phi} \times \prod_{j=1}^{N_{\lambda}}P_{j}^{F_{\ell}},
\end{equation}
where $\calD$ represents the measured data, $\bm{\Theta}$ represents all the model parameters.
$N_{\rm RM}$ is number of RM observations, $N_{\rm G}$ is the number of GRAVITY observations 
(all the baselines) and $N_{\lambda}$ is the corresponding number of wavelength bins.
In light of Bayes' theorem, the posterior probability distribution for $\bm{\Theta}$ is given by
$P(\bm{\Theta} | \calD)=P(\bm{\Theta}) P(\calD | \boldsymbol{\Theta})/P(\calD)$,
where $P(\bm{\Theta})$ is the prior distribution of the model parameters and $P(\calD)$ is a 
normalization factor.

\begin{figure*}\label{fig:fitting-results}
\vglue -0.5cm
\begin{center}\includegraphics[scale = 0.45]{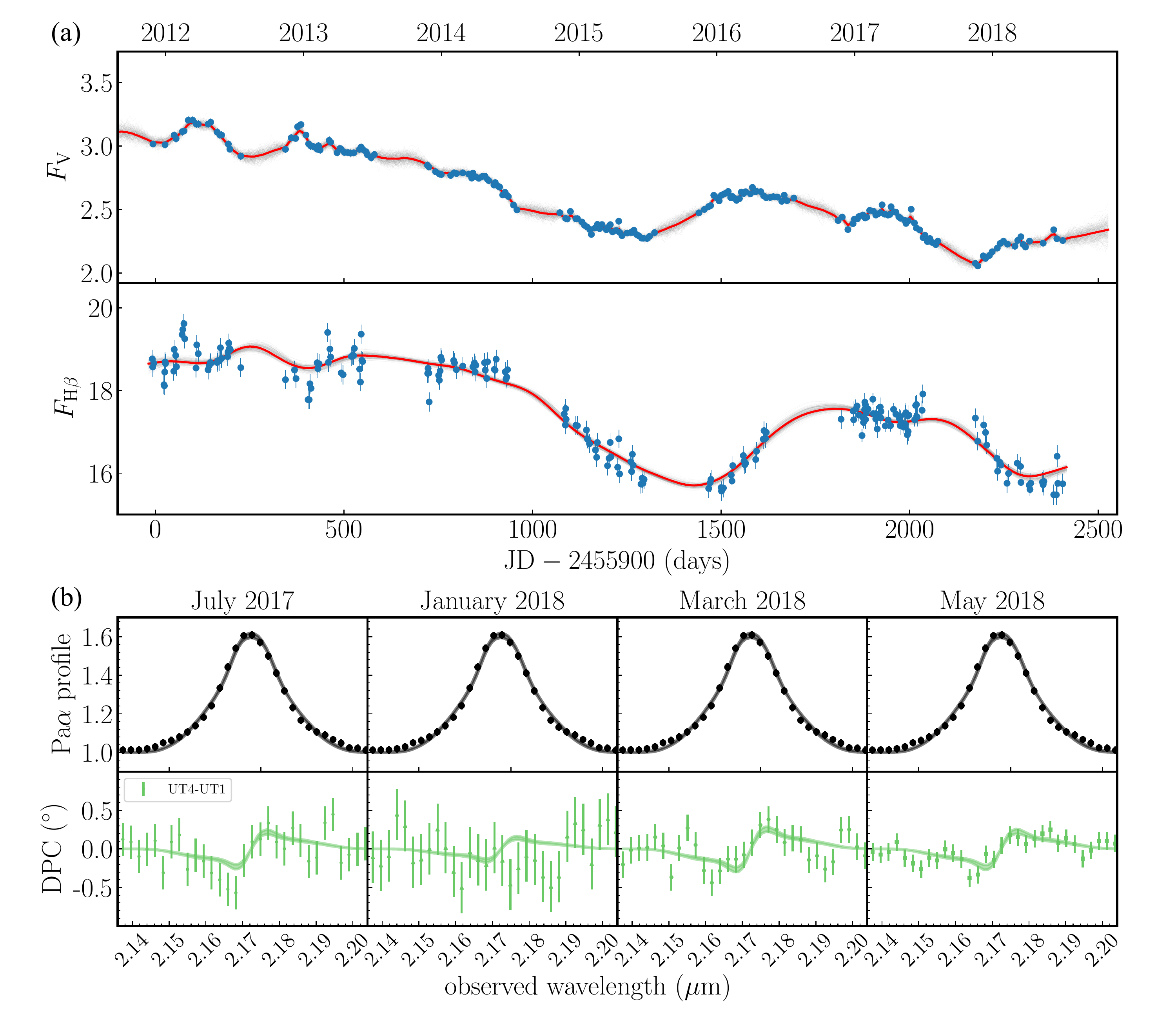}\end{center}
\vglue -1cm
    {
    \footnotesize {\bf Figure 1 $ |$ Joint fittings of RM and GRAVITY observations.}
    Panel {\it a}: one-dimensional fitting of the RM data since 2012 (avoiding 
    contaminations of the relativistic jet, see details in Methods) through the BLR model.
    Blue points are data points with $1\,\sigma$ error bars; red lines are the best fitting
    results of LCs; gray ones are the results using 200 groups of model parameters randomly
    drawn from their probability distribution.
    The H$\beta$ fluxes are in units of $10^{-13}\,{\rm erg\, s^{-1}\, cm^{-2}}$, and $F_{V}$
    in $10^{-14}\,{\rm erg\,s^{-1}\,cm^{-2}\,\AA^{-1}}$ converted from $V$-band magnitudes.
    The scatter of the H$\beta$ LC around the beginning 
    of 2012, 2013 significantly contributes to $\chi^{2}=1.61$, which is relatively larger than that 
    of GRAVITY data. Panel {\it b}: Fittings of the differential phase curves (DPC; green color points) of
    the baseline UT4-1 and Pa$\alpha$ line profiles (black points; $\chi^{2}=1.33$) as an example, and the 
    complete fittings of all baselines are given in Methods. All the data points are with $1\,\sigma$ 
    error bars; thick lines are the best fitting through the model with values of 
    parameters given in Table 1; translucent thin lines are fitting results 
    using 200 groups of model parameters randomly drawn from their probability distribution.
}
\end{figure*}

In the Methods section, we show evidence for jet contamination of the observed continuum 
giving rise to trending effects before 2012 (see details in Ref.\cite{Li2019}).
Only the RM data after 2012 are thus taken into account in the joint analysis. The best-fittings 
to the SARM data are shown in Figure 1, \ref{fig:fitting-results}, 
and the projection of probability density distributions are shown in Figure 2 
for the three key parameters ($R_{\rm BLR}, \bhm, D_{\rm A}$). See Supplementary Figures 2 and 3
for the fittings of whole differential phase curves and projection of probability 
density distributions of the complete parameters.
The median value and $1\sigma$ error bar for each parameter are also given there and listed in 
Table 1. Some parameters have values similar to those in Ref.\cite{Sturm2018} 
within error bars, or different within reasonable ranges. The present joint analysis 
generates an angular distance of
$D_{\rm A}=551.5_{-78.7}^{+97.3}\,{\rm Mpc}$
with a relative statistical error of $|\Delta D_{\rm A}|/D_{\rm A}\approx 0.16$ 
on average. This is a very encouraging accuracy for the joint analysis of the first SARM data,
demonstrating the power of the present analysis as a feasible tool for measuring extragalactic 
distances. {
The SARM measurements as a geometrical method 
avoid various calibrations and corrections used in the popular measurements through Cepheid 
variable stars\cite{Freedman2017} and type Ia supernovae (SNIa)\cite{Riess1998,Perlmutter1999},
such as extinction corrections in both tools, necessary calibrations through
the standardization and cosmic ladders in the latter.} 

Systematic errors of results in the joint analysis are mainly governed by three factors.
As the first step of the SARM approach, we use the simplest model of the BLR 
to simultaneously fit GRAVITY data and RM LCs, namely, for one dimension fitting 
(2D model will include H$\beta$ profile and its variations; see Methods for more
explanations). Since Pa$\alpha$ and H$\beta$ lines are both from $n=4$ energy 
level to $n=3,2$, respectively, in principle, GRAVITY-measured Pa$\alpha$ regions should 
share the same regions with H$\beta$ line measured by RM. For the current case of 3C 273, 
however, they show small difference of the sizes ($\sim 13\%$) likely due to optical depths 
of the two lines in light of their profile width (see the Methods for details). Fortunately, 
the SARM analysis can completely avoid this problem if observations are for the same line, 
i.e., mapping Pa$\alpha$ in near infrared bands (such an RM campaign is actually in planning). 
Second, lengths of GRAVITY observations and RM campaigns
are quite different and they measure the variable part and entire regions, respectively. 
This may give rise to differences measured by the two observations, however,
we can conveniently justify this by comparing the RMS and mean spectra. We found that
the two spectra of 3C 273 are similar in widths and shapes\cite{Zhang2019}, implying that 
GRAVITY-measured regions are about identical to the RM-measured ones. Moreover, the dynamical 
timescale of the BLR is much longer than the length of our RM campaigns. The two conditions 
guarantee the validity of the joint analysis, and such a kind of
systematic errors can be minimized. Third, non-disc like geometry or radial motion of BLR clouds 
could result in systematic errors in the analysis. In practice, fortunately, the velocity-resolved 
delays or its 2D delay maps from the maximum entropy method\cite{Horne1991} provide key information 
to justify deviation of geometry and kinematics of the BLR from the simplest model, such as 
a disc-like geometry of the BLR with Keplerian rotation in 3C 273 supported by
both velocity-resolved delays\cite{Zhang2019}
and interferometric data\cite{Sturm2018}. In principle, 
all these factors can be maximally avoided or at least observationally tested to
reliably get systematic errors of measurements (see details of observational strategies
in the Methods).

We can measure the Hubble constant $H_{0}$ through the SARM-based distance of 3C 273.
Employing the $z-D_{\rm A}$ relation\cite{Peacock1999}, we have
$H_{0}=71.5_{-10.6}^{+11.9}\,{\rm km\,s^{-1}\,Mpc^{-1}}$
for a cosmology of $\Omega_{\rm M}=0.315$ and $\Omega_{\Lambda}=0.685$ determined  
by the Planck CMB measurements\cite{Aghanim2018} (but $H_{0}$ very weakly depends 
on $\Omega_{\rm M}$ and $\Omega_{\Lambda}$ for the current case). Considering the current 
accuracy of distance measurements of 3C 273 (this is mainly controlled by the error 
bars of the DFC measured by GRAVITY, which can be improved significantly in future 
observations), we have uncertainties of $H_{0}$ given by 
$|\Delta H_{0}|/H_{0}=|\Delta D_{\rm A}|/D_{\rm A}\lesssim 3\,N_{30}^{-1/2}$ per 
cent for a AGN sample of a reasonable size, where $N_{30}=N/30$ is the number of AGNs. A 
selection is done for $K$-band brighter than $\lesssim 11.5$mag in the Supplementary Information
(the $\sim 13\%$ uncertainties are not included since it can be in principle eliminated). 
Such a precision is enlightening for a test of the current $H_{0}$-tension\cite{Riess2019}.
Targets of a future SARM project should be focused on AGNs with smooth and symmetric 
broad emission-line profiles in order to reduce systematic errors (or GRAVITY$^{+}$ as a next 
generation of GRAVITY is designed for fainter targets in the near future, making target 
selection much easier). Advantages of the SARM-based measurements are obvious. First, the 
distance measurements are geometrical for $H_{0}$.
Though the SARM-based measurements depend on physical models, they can be observationally 
tested in advance. Second, SARM targets can be easily selected from existing AGN catalogs 
(and can be more distant than Cepheids and SNIa). Repeat measurements  
(invoking multiple campaigns of 2-4m class telescopes to simultaneously 
monitor the targets) allow us to test and greatly reduce systematic error bars. Third, the number of 
targets spatially distributed over the sky allows us to obtain high precision measurements 
of the $z-D_{\rm A}$ relation for different directions in order to test the potential anisotropy of 
the accelerating expansion of the Universe\cite{Cai2013} and advance the understanding of cosmological 
physics\cite{Weinberg2008}.


\begin{table*}\label{tab:BLR}
\footnotesize
{\centering
\caption{Parameters used in the BLR model and the SARM results of 3C 273\label{tab:parameters}}
\vglue 0.2cm
\begin{tabular}{lllcll}\hline\hline
	Parameters              & meanings                          & GRAVITY & RM$^{\rm 1D}$  &Joint analysis & Prior ranges \\ \hline
$\mathscr{F}$               & fractional inner radius of the BLR& $\surd$\, ($0.23\pm 0.08$)& $\surd$& $0.49_{-0.20}^{+0.12}$ & $[0,1]$ \\
$\beta$                     & radial distribution of BLR clouds & $\surd$\, ($1.4\pm0.2$)   & $\surd$& $1.09_{-0.40}^{+0.91}$ & $[0,4]$  \\
$\theta_{\rm opn}(^{\circ})$& half opening angle of the BLR     & $\surd$\, ($45_{-6}^{+9}$)& $\surd$& $39.96_{-3.72}^{+4.01}$ &  $[0,90]$\\ 
    $i_{0}(^{\circ})$       & inclination angle of the BLR      & $\surd$\, ($12\pm2$)      & $\surd$& $8.41_{-0.91}^{+0.99}$ & $[0,90]$\\
PA($^{\circ}$)              & position angles                   & $\surd$\, ($210_{-9}^{+6}$)& & $210.99_{-4.63}^{+3.67}$ & $[0, 520]$\\    
$R_{\rm BLR}$(ltd)          & averaged linear sizes & & $\surd$  & $184.17_{-8.57}^{+16.77}$ & $[1,10^3]$\\
$\bhm(10^{8}M_{\odot})$     & supermassive black hole mass      & $2.6\pm 1.1$ & & $5.78_{-0.88}^{+1.11}$ & $[10^{-2}, 10]$\\
    $D_{\rm A}$(Mpc)        & absolute angular distance         &  $550$ (assumed)& & $551.50_{-78.71}^{+97.31}$ & $[10,10^4]$\\ 
\hline
$\xiBLR$\,($\mu$as)         & averaged angular sizes & $\surd$ ($46\pm 10$) &  & $59.70^{+8.72}_{-10.31}$  \\
$\zeta\,(10^{-2})$ & dimensionless velocity parameter & $\surd$ ($1.01\pm 0.22$) &  & $1.34_{-0.06}^{+0.12}$ &\\
\hline
\end{tabular}}
\vglue 0.2cm
{\footnotesize\,   \,\,  
Notes: ``$\surd$'' means that the parameter can be determined by GRAVITY or RM data. 
Numbers in brackets behind ``$\surd$'' are median values with uncertainties of 90\% from fittings 
of GRAVITY data\cite{Sturm2018} for a convenient comparison. Values determined by the joint 
analysis are medians of the posterior distributions with uncertainties of 68\% confidence ranges.
RM$^{\rm 1D}$: one-dimensional reverberation mapping (RM), in which only flux variations of broad 
emission lines are fitted. $\xiBLR=\RBLR/D_{\rm A}$ (the angular sizes) and 
$\zeta=(GM_{\bullet}/R_{\rm BLR})^{1/2}c^{-1}$ are reduced quantities for the fitting.
 }
\end{table*}



\clearpage
 
\begin{addendum}

\item {We are grateful to three anonymous referees for useful reports improving this paper.
We acknowledge the support by National Key R\&D Program of 
China through grant - 2016YFA0400701, by NSFC through grants NSFC-11991050, -11873048, -11833008, 
-11573026, and by Grant No. QYZDJ-SSW-SLH007 from the Key Research Program
of Frontier Sciences, CAS, by the Strategic Priority Research Program of the Chinese 
Academy of Sciences grant No.XDB23010400. E. Sturm is thanked for useful information 
of GRAVITY and the future GRAVITY$^{+}$ capabilities. JMW is grateful to M. Brotherton 
for careful reading the manuscript and useful discussions, to Bo-Wei Jiang, Dong-Wei Bao, 
Wei-Jian Guo and Sha-Sha Li who helped for the target selections.
}

\item[Author Contributions] 
JMW conceived this project and wrote the paper. YYS, YRL and JMW made all calculations. 
ZXZ and PD made observations and perform data reduction. JMW, ZXZ and PD selected targets of the
future SARM projects. All the authors discussed the contents of the paper.

\item[Correspondence] 
Correspondence and requests for materials
should be addressed to Jian-Min Wang (email: wangjm@ihep.ac.cn).
Jian-Min Wang: https://orcid.org/0000-0001-9449-9268\\

\item[Competing Interests] The authors declare that they have no
competing financial interests.

\end{addendum}

\clearpage

\begin{figure*}
\begin{center}\includegraphics[scale = 0.6]{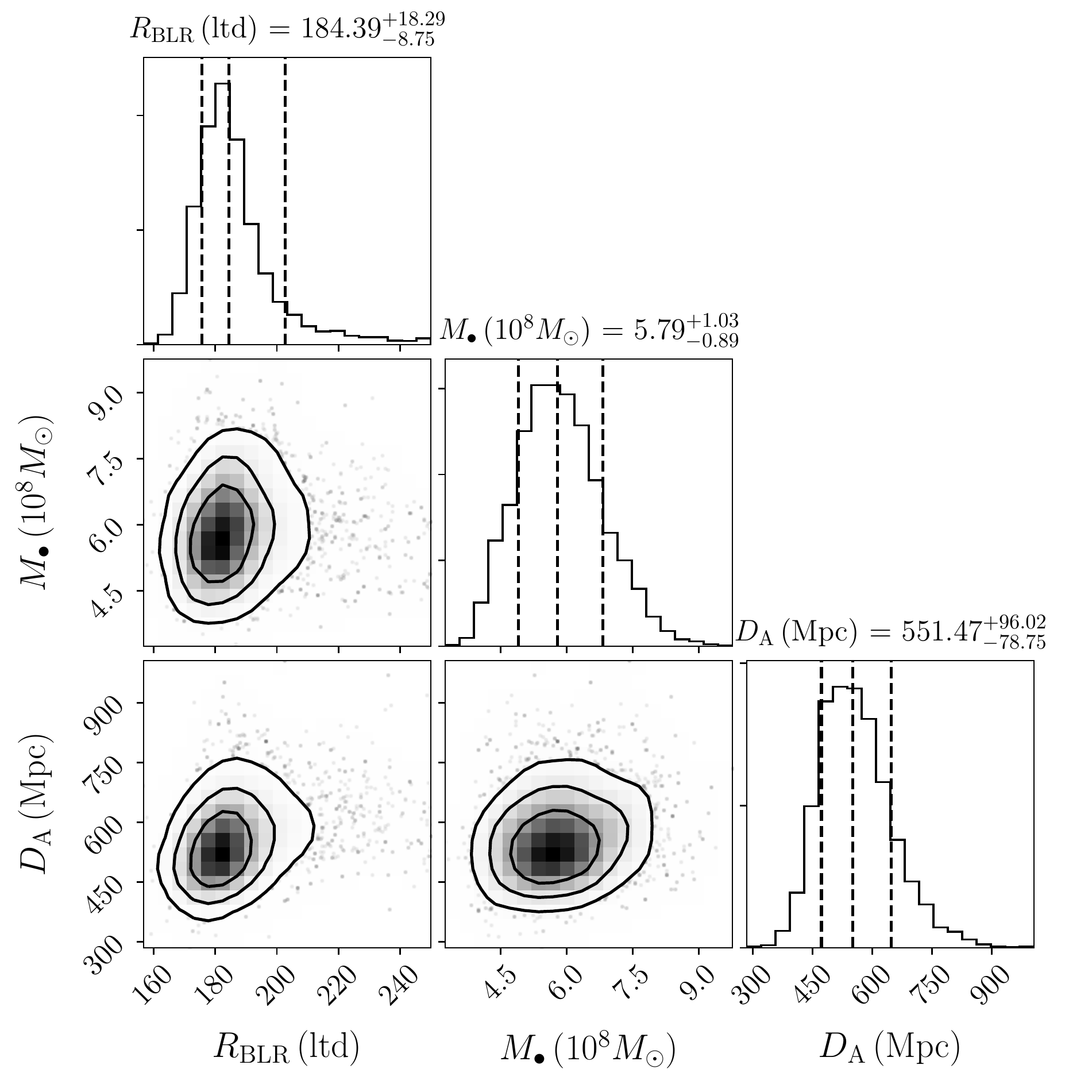}\end{center}
    {\footnotesize{\bf Figure 2 $ |$ Results of black hole mass and distances.}
    Probability density distributions of three key parameters of the BLR and 
    angular distances based on the joint analysis. The best values of the parameters are 
    given on the tops of panels. Error bars are quoted 
    at $1\sigma$ level, which are given by each distributions.
    The dashed lines in the one-dimensional distributions are the $16\%$, $50\%$ and $84\%$ quantiles, 
    and contours are at $1\sigma$, $1.5\sigma$ and $2\sigma$, respectively.
}
\label{fig:contour3}
\end{figure*}

\clearpage

\begin{methods}

\subsection{Spectroastrometry.}
``Differential Speckle Interferometry" as the progenitor of the spectroastrometry was 
first suggested by J. M. Becker\cite{Becker1982} and its feasibility was demonstrated 
by R. Petrov\cite{Petrov1992}. We follow the description of interferometry in Ref.\cite{Petrov1989}
(but see also Ref.\cite{Rakshit2015} for more extensive discussions). Spectroastrometry 
is a powerful tool of high spatial resolution. Given the surface brightness distribution 
of the regions, we have
\begin{equation}\label{eq:ph-center}
\bm{\epsilon}(\lambda) = 
\frac{\int \bm{\alpha} \calO(\bm{\alpha},\lambda)  \dd[2]{\bm{\alpha}}}
{\int \calO(\bm{\alpha},\lambda) \dd[2]{\bm{\alpha}}},
\end{equation}
where $\calO(\bm{\alpha},\lambda)=\calO_{\ell}+\calO_{\rm c}$ is the surface brightness 
distribution of the source contributed by the BLR and continuum regions, respectively, and
$\bm{\alpha}$ is the angular displacement on the celestial sphere. Given the geometry and 
kinematics of a BLR, its $\calO_{\ell}$ can be calculated for one broad emission line 
with the observed central wavelength $\lambda_{\rm cen}$ through
\begin{equation}\label{eq:calO}
\calO_{\ell}=\int \frac{\Xi_r F_{\rm c}}{4 \pi r^2} f(\bm{r},\bm{V})
\delta\!\left(\bm{\alpha}-\bm{\alpha}' \right) 
          \delta\!\left(\lambda -\lambda^{\prime}\right)
         \dd[3]{\bm{r}} \dd[3]{\bm{V}},
\end{equation}
where $\lambda^{\prime}=\lambda_{\rm cen}\gamma_{0}\left(1+\bm{V}\vdot\nobs/c\right)
\left(1-R_{\rm S}/r\right)^{-1/2}$ includes gravitational shifts due to the central 
black hole, $\gamma_{0}=\left(1-V^2/c^2\right)^{-1/2}$ is the Lorentz factor, 
${\bm \alpha}^{\prime}=\left[\bm{r}-\left(\bm{r}\vdot\bmn\right)\bmn\right]/D_{\rm A}$,
$\bm{r}$ is the displacement to the central BH, $\Xi_r$ is the reprocessing coefficient 
at position $\bm{r}$, $f(\bm{r},\bm{V})$ is the velocity distribution of the clouds 
at that point, $F_{\rm c}$ is ionizing fluxes received by an observer, and
$\bmn=(0,\sin i_{0},\cos i_{0})$ 
 is the unit vector pointing from the observer to the source.
Introducing the fraction of the emission line to total ($\fline$), we have
\begin{equation}\label{eq:epsilon}
\bm{\epsilon}(\lambda) = \fline\,\bm{\epsilon}_{\ell}(\lambda),
\end{equation}
where
\begin{equation}
\bm{\epsilon}_{\ell}(\lambda) = \frac{\int \bm{r} \calO_{\ell}  
   \dd[2]{\bm{\alpha}}}{\int \calO_{\ell} \dd[2]{\bm{\alpha}}},\,\,\,\fline 
= \frac{\Fline(\lambda)}{F_{\rm tot}(\lambda)},\, \ \ \Fline(\lambda) 
= \int \calO_{\ell} \dd[2]{\bm{\alpha}}, \ \ 
   F_{\rm tot}(\lambda)=\Fline(\lambda) + F_{\rm c}(\lambda). \nonumber
\end{equation}
Inserting Equations (\ref{eq:epsilon}), (\ref{eq:calO}) and (\ref{eq:ph-center}) into (\ref{eq:phase}), 
we can obtain phase curves.
Since $\bm{B}/\lambda\sim 100{\rm m}/2.2\mu{\rm m}$ and $\bm{\epsilon}\sim 100{\mu\rm as}$,
$\phi_{*}$-amplitudes are expected to be at a level of a few degrees for spatial resolution 
of compact objects. If the BLR model is specified, the spectroastrometric technique is able to 
efficiently improve the spatial resolution. 

\subsection{Reverberation mapping.}
AGNs and quasars are radiating with huge power and their spectra are prominently characterized 
by broad emission lines from NIR, optical to ultraviolet 
bands\cite{Osterbrock1986,Osterbrock1989,Ho2008}.
The standard model of AGNs is accretion onto supermassive black holes (SMBHs) located in galactic 
centers producing powerful radiation\cite{Rees1984}. Emission lines from the photoionized gas 
are broadened by fast motions under the gravitational potential of the SMBHs and appear with a
full width at half maximum (FWHM) spanning from $\sim 10^{3}$ to a few $10^{4}\kms$. According to 
energy conservation, the covering factor of the BLR clouds is about 10\%, representing a fraction 
of reprocessing energy released by the accretion. As a natural consequence of 
photoionization, the emission lines will follow variations of the continuum, but with a 
delay denoted as $\tau_{\rm BLR}$. This delayed response is known as the reverberation 
of the BLR\cite{Peterson1993}. Considering that the recombination timescale 
$\tau_{\rm rec}\approx (n_{e}\alpha_{\rm B})^{-1}\approx 0.1\,n_{10}^{-1}\,$\,hr is much
shorter than $\tau_{\rm BLR}$, the delays of the emission lines represent the linear 
dimension of the emission line regions, where $n_{10}=n_{e}/10^{10}{\rm cm^{-3}}$ is 
electron density of clouds and $\alpha_{\rm B}$ is the case B recombination 
coefficient\cite{Osterbrock1989}. 

RM observations measure the LCs of 
broad emission lines and continuum, and then allow us to investigate the temporal relation between 
the lines and the continuum for the BLR geometry and kinematics\cite{Peterson1993}. 
Echo of emission lines to the continuum was suggested earlier\cite{Bahcall1972}, but observation
campaigns began to measure it since 1980s. Nowadays, RM technique is regarded as the most powerful 
tool of measuring the central black hole mass\cite{Peterson2014a} in studies of the so-called
coevolution of SMBHs and galaxies\cite{Ferrarese2005}. There are $\sim$ 100 AGNs with robust 
H$\beta$ RM measurements (summarized in Ref.\cite{Kaspi2017}), this number is expected to 
dramatically increase contributed by several RM groups over the word in near future.
In practice, $\tau_{\rm BLR}$ can be easily measured from the simple cross-correlation 
function, however, its exact meaning can only be specified through modeling the BLR.

\subsection{Parameterized BLR.}
The understanding of BLRs has been advanced much after great efforts of RM campaigns for about 
100 AGNs during the last several decades\cite{Peterson1998,Kaspi2000,Bentz2013,Du2018}. 
It has been found that, except for optical \feii-strong AGNs\cite{Du2018,Du2019}, BLR sizes 
follow a well-established $R-L$ 
relation\cite{Bentz2013,Du2018} agreeing with the consequence of photoionization of isotropic 
ionizing sources. These \feii-strong AGNs\cite{Du2018} are mostly super-Eddington objects powered 
by slim accretion discs\cite{Abramowicz1988,Wang2014b, Wang2014a}, breaking the assumption of 
isotropic ionizing sources made in the explanation of H$\beta$ line reverberation\cite{Wang2014a}. 
Self-shadowing effects due to the puffed-up inner part of slim disks greatly obscure illuminations
of the BLR so that H$\beta$ lags are significantly shortened\cite{Wang2014a,Du2018}.
For \feii-weak AGNs, they are likely powered by geometrically thin accretion discs and can be well 
approximated as isotropic sources. Secondly, a disc-like BLR has been generally
found in many broad-line Seyfert 1 galaxies from velocity-resolved delay 
maps\cite{Grier2013,Bentz2013,Lu2016,Du2016b,Xiao2018}, even in some narrow-line Seyfert 
1 galaxies\cite{Du2016b}. Moreover, the differential phase curves of the Pa$\alpha$ 
line in 3C 273 directly show evidence for a Keplerian rotation of flattened
disc as the BLR. Third, there is growing evidence 
for Keplerian rotation of the BLR clouds through multiple campaigns of several AGNs, such 
as, NGC 5548, NGC 3783, 3C 390.3 and NGC 7469\cite{Peterson2004}, stratified 
radial structure of the BLR according to the ionization energy of ions in 
NGC 5548\cite{Peterson2000}, or vertical structure of the 
BLR (see Figure\,24 in Ref.\cite{Kollatschny2014}). Additionally, the angle between the 
direction of orbital angular momentum and the $Z$-axis is uniformly distributed 
over [0, $\theta_{\rm opn}$], which is used in BLR modeling\cite{Sturm2018}.
As the zero-order approximation of the present scheme, we simply assume that the BLR 
clouds are orbiting with Keplerian velocity around the central black hole in this paper. 
In such a characterized BLR, emission line profiles are usually symmetric.
See Supplementary Figure 1 for the BLR geometry.

With the goal of illustrating a new scheme to determine distances, we employ a stream 
lined model of the BLR, rather than a comprehensive BLR model 
(with about $\sim 20$ parameters) pursuing fine fitting of the observational data 
as done in Ref.\cite{Pancoast2011,Pancoast2014,Li2013,Li2018,Williams2018}. 
The BLR model could include more components in the future if GRAVITY data is significantly 
improved. 

We point out that the ``clouds'' used in this paper can be generally understood
as elements of the BLR if it is a kind of fluid. These clouds could be supplied by the 
central black hole tidal capture of clumps from torus\cite{Wang2017} or winds from accretion
disc\cite{Czerny2011}. The former model keeps a quasi-stationary state of the BLR whereas some
clouds switch to the accretion disc of the black hole.

\subsection{RM$^{\rm 1D}$ modeling.}
Detailed descriptions of RM$^{\rm 1D}$ modeling of the BLR are given by Ref.\cite{Li2013,Li2018}. 
Here we briefly summarize the necessary formulations for the reader's convenience. In order to 
interpolate and extrapolate the sampled LC of the varying continuum, we use the damped random 
walk (DRW) model to describe the continuum variations\cite{Kelly2009,Zu2013}. For a time series 
of $\bmy$, the measured data can be expressed by
$\bmy=\bms+\bmnn+\bmE q$,
where $\bms$ is the variation signal that is described by the DRW model, $\bmnn$ is the 
measurement errors, $q$ is the mean values of the series, and $\bmE$ is a vector with all 
unity elements. The covariance function of the DRW model is given by
\begin{equation}\label{eq:S}
S(t_1,t_2)=\sigmad^{2}\exp\left(-\frac{|t_{1}-t_{2}|}{\taud}\right),
\end{equation}
for any two points at times $t_{1}$ and $t_{2}$, where $\sigmad$ is the long-term standard 
deviation of the variations and $\taud$ is the typical timescale of the
variations\cite{Kelly2009}. Supposing that both $\bms$ and $\bmnn$ are Gaussian and
unrelated\cite{Li2013,Li2018}, the best estimate of $q$ is given by
\begin{equation}
\hat{q}=\frac{\bmE^{T}\bmC^{-1}\bmy}{\bmE^{T}\bmC^{-1}\bmE},
\end{equation}
where the superscript ``{\it T}\,'' denotes the transposition, $\bmC=\bmS+\bmN$, $\bmS$ 
and $\bmN$ are the covariance matrix of the signal $\bms$ and noise. Using Bayes' theorem, 
we can recover the damped random walk process to determine the best values of $\sigmad$ and 
$\taud$ for a given set of the series. The most probable estimate of the variation signal 
$\bms$ at any time $t_{\star}$ is given by
\begin{equation}
\hat{\bms}=\bmS^{T}\bmC^{-1}(\bmy-\bmE\hat{q}).
\end{equation}
A typical realization for the continuum LC is\cite{Li2018}
\begin{equation}\label{eq:fc}
f_{\rm c} = {({\boldsymbol u}_s + \hat\bms)} + \bmE(u_q-\hat q),
\end{equation}
where ${\boldsymbol u}_s$ follows a Gaussian  process with a zero mean and covariance of 
${\boldsymbol Q}=[\bmS^{-1}+\bmN^{-1}]^{-1}$, and $u_q$ follows a Gaussian process with 
a zero mean and covariance of $(\bmE^{T}\bmC^{-1}\bmE)^{-1}$. We treat ${\boldsymbol u}_s$
and $u_q$ as free parameters, which are further constrained by the LC data of the emission 
line.

All calculations are done using the coordinates shown in Supplementary Figure 1. 
Given the BLR geometry and kinematics, we can calculate the
response of the entire BLR to the varying continuum in order to fit the observed data.
The time-dependent fluxes of the broad emission line can be calculated by summing up
the reprocessing emissions from all the BLR clouds as
\begin{equation}\label{eq:fl}
f_{\ell}(t)=\int\dd{\bm{r}}\dd{t^{\prime}}\frac{\Xi_{r}f_{\rm c}(t^{\prime})}{4\pi r^{2}}n(\bm{r}) 
\delta\left(t^{\prime}-t+\tau\right)
\end{equation}
where $\tau = (r - \bm{r}\vdot\bm{n}_{\rm obs})/c$, $\Xi_{r}$ is the reprocessing 
coefficient and $n(\bm{r})$ is the number density of the clouds. Given the BLR model, 
Equations (\ref{eq:fc},\ref{eq:fl}) are used to fit the observed LCs denoted as 
the RM$^{\rm 1D}$ analysis.

\subsection{Observational data and de-trending.}
We note that the varying continuum has a long decreasing trend, but H$\beta$ 
LC does not have the same trend\cite{Zhang2019} (also see Supplementary 
Figures 4 and 5). 
It has been suggested 
that de-trending is an efficient way to improve the RM analysis of long-term
secular variability of continuum\cite{Welsh1999,Peterson2004}. In the present scheme
of fittings, we take a linear form of de-trending as $f_{\rm c}\propto k_{\rm c}(t-t_{0})$ 
for the continuum LCs, where $k_{\rm c}$ is determined by the joint analysis.

\subsection{\it Contamination.}
In Supplementary Figure 4, 
we show the $\gamma$-ray LC monitored by the LAT of the 
{\it Fermi} satellite (https://fermi.gsfc.nasa.gov/ssc/data/access/). The 
$\ge 30$GeV emission must be generated by the most inner part of the jet\cite{Blandford1995}, 
and thus contamination could be tested by $\gamma$-rays. The {\it Fermi} monitoring campaign
is continuous without season gaps, allowing us to test the long-term trend. We generate the 
$\gamma$-ray LC according to the {\it Fermi} user's Guider (see also the $\gamma$-ray 
LC in Ref.\cite{Meyer2019}). It is obvious that there is a giant $\gamma$-ray flare in the second 
half year of 2009 and first months of 2010, and the flare has a long tail lasting to the 
end of 2011. It is not the goal of the present paper to make a detailed comparison between the
$\gamma$-ray and $V$-band LCs, but it is quite obvious that the flare significantly contributes
to the observed optical continuum. It is also clear that the response of the broad H$\beta$ 
line to the optical continuum is weak. The jet contamination has been investigated\cite{Li2019} 
through analysis of multiple long-term light curves from radio to optical band, as well as 
polarization. It is clear that the long term trend of 5100\AA\, continuum is due to the jet 
contaminations. See details in this reference.

\subsection{\it Weak response of the broad H$\beta$ line.}
Entire LCs are shown by Supplementary Figure 5{\it a}, 5{\it b} and 5{\it c}.
With the de-trended continuum LC, we obtained a lag of 
$\tau_{\rm BLR}=146\pm 8$\,days\cite{Zhang2019}.  We shifted the H$\beta$ LC backward by the 
$(1+z)\tau_{\rm BLR}$ and multiplied $F_{\rm H\beta}$ by a reasonable factor for a compare 
with the de-trended continuum in Supplementary Figure\, 5{\it d}. 
It is clear that the response of H$\beta$ is weak before 2012. One plausible reason is that 
the optical continuum is contaminated (UV is less significantly) by the jet located 
outside the BLR, but the BLR has no response to this continuum component. We only focus 
on the linear response at the zero-order of the model, rather than non-linear responses 
of the BLR clouds in this paper.

\subsection{Fittings.}
Clouds are randomized to distribute along a given orbit. We use this prescription 
of cloud distribution for the calculation of transfer functions, line profiles, and 
differential phase curves. Special and general relativistic effects are included.
We take the priors of the BLR parameters in sufficiently wide ranges in order to 
guarantee the unique solutions of the model. Ranges of parameters are provided in 
Table 1. The priors of 
$(\mathscr{F},\beta,\theta_{\rm opn},i_0,{\rm PA})$ are uniform over the given 
intervals, while those of $(R_{\rm BLR},M_{\bullet},D_{\rm A})$ are uniform in 
log scale. The cosmological dilation factor of $(1+z)$ has been included in generating 
LC to fit observations of 3C 273 in the BLR modeling. Using the Diffusive Nested 
Sampling\cite{Brewer2018}, we obtain a total of $5000$ 
samples for all model parameters. 
We generate probabilities in Equation (\ref{eq:probability}), $P^{f_{\ell}}_{i}$ 
for H$\beta$ reverberation, $P_{i,j}^{\phi}$ for differential phase curves and 
$P_{j}^{F_{\ell}}$ for profiles of the Pa$\alpha$ line, 
\begin{equation}
P^{f_{\ell}}_{i}=\frac{1}{\sqrt{2 \pi\sigma_{\ell}^{2}}} 
\exp \left\{-\frac{\left[f_{\ell,i}^{\rm obs}-f_{\ell,i}^{\rm mod}\left(f_{\rm c, obs} | 
\bm{\Theta}\right)\right]^{2}}{2\sigma_{\ell}^{2}}\right\},
\end{equation}

\begin{equation}
P_{i,j}^{\phi}=\frac{1}{\sqrt{2 \pi \sigma_{\phi_{ij}}^{2}}} 
\exp \left\{-\frac{\left[\phi_{i,j}^{\rm obs}-\phi_{i,j}^{\rm mod} 
\left(\bm{\Theta}\right)\right]^{2}}{2\sigma_{\phi_{ij}}^{2}}\right\},
\end{equation}

\begin{equation}
P_{j}^{F_{\ell}}=\frac{1}{\sqrt{2 \pi \sigma_{\rm F}^{2}}} 
\exp \left\{- \frac{\left[F_{\ell,j}^{\rm obs} - F_{\ell,j}^{\rm mod}\left(\bm{\Theta}\right)\right]^{2}}
{2\sigma_{\rm F}^{2}}\right\},
\end{equation}
where $f_{\ell,i}^{\rm obs}$, $F_{\ell, j}^{\rm obs}$, and $\phi_{i,j}^{\rm obs}$ are the 
observed line flux, line profile, and interferometric phase of the emission line with 
measurement uncertainties $\sigma_{\ell}$, $\sigma_{\phi_{ij}}$, and $\sigma_{\rm F}$, 
respectively, and $(f_{\ell,i}^{\rm mod},F_{\ell,j}^{\rm mod},\phi_{\ell,i}^{\rm mod})$ 
are the corresponding predicted values from the BLR model.

The joint analysis of the SARM data shows the reduced $\chi^{2}_{\rm G}=1.33$ for GRAVITY data
and $\chi^{2}_{\rm RM}=1.61$ for the RM data. The $\chi^{2}_{\rm RM}$ is a little bit higher 
than $\chi^{2}_{\rm G}$, but note that this could be caused by a couple of points in the H$\beta$ 
LC significantly deviating from the model. Excluding these points, we find that the $\chi^{2}$ will 
be greatly reduced. Considering the major goals of the present paper, we keep this fitting 
with the $\chi^{2}=1.61$ as resultant fittings. We found $k_{\rm c}=(7.9\pm 0.2)\times 10^{-5}$ 
from the joint fitting, which is consistent with the results in Ref.\cite{Li2019}

We generated mock data to test the present scheme. Error bars of GRAVITY phase curves are
set to be $\sim 25\%$ and RM data at the level of the 3C 273 campaign. We find that the 
generated parameters of the model from the mock data are in good agreement with the input 
within 10\%. This demonstrates that the present joint analysis is feasible for simultaneous
determinations of distances and black hole mass of AGNs.

\subsection{Quasars as cosmological probes.}
Quasars are the most luminous and long-lived celestial objects in the Universe. After their 
discovery, they were instantly suggested as probes for 
cosmology\cite{Sandage1965,Hoyle1966,Longair1967,Baldwin1977} by arguing that some properties
of quasars can serve as ``standards''. However, these efforts were 
not successful because of poor understanding of quasar physics. Recently, interests in 
applications of quasars to cosmology have arisen again by selecting special individual
objects or populations in light of their well-understood properties.

Direct measurements of distances through very long base interferometry (VLBI) observations of 
water masers in NGC 4258 ($z\approx 0.0015$)\cite{Humphreys2013}, and torus diameters through 
direct imaging observations of NGC 4151 ($z\approx 0.0033$)\cite{Honig2014} have been suggested
for cosmology,
but these methods are limited either by the rare sources of water masers or the long period of 
NIR monitoring campaigns. Parallax of quasar BLR was suggested for measurements of 
distances\cite{Elvis2002,Quercellini2009}, 
which needs an interferometer with a baseline of $\sim 100$km, but GRAVITY at The Very Large Telescope
Interferometer (with baselines of only $\sim 100$m) employs spectroastrometry to efficiently reduce 
the required baseline for 10$\mu$as resolution. Super-Eddington accreting massive black holes 
(SEAMBHs over much wider ranges of redshifts) from their saturated 
luminosity\cite{Abramowicz1988,Wang1999,Wang2013,Du2014,Wang2014a,Marziani2014,Du2018} were suggested 
for cosmology and show potential feasibilities of application to the high-$z$ Universe\cite{Cai2018}. 
Moreover, the following relations have been also suggested for cosmology, such as the
well-known $R-L$ relation of sub-Eddington AGNs\cite{Watson2011,Czerny2013,King2015} or
its improved version\cite{Martinez2019}, NIR continuum reverberation correlation\cite{Yoshii2014},
X-ray variance versus luminosity\cite{LaFranca2014}, and the non-linear UV versus X-ray 
luminosity relation\cite{Risaliti2019}. Quasars as cosmological 
objects seem to be a promising tool to probe the Universe in future\cite{Czerny2018,Marziani2019},
but much work needs to be done for precision cosmology as a robust probe. 

We would like to point out that all the methods mentioned above need calibrations of cosmic 
ladders except for the way of the VLBI water maser and NIR RM techniques. 
The application of the $R-L$ relation to cosmology has been 
initiated\cite{King2015,Czerny2019,Hoormann2019}, but emission-line lags for high-$z$ quasars 
are hard to measure because of the dilation factor of $(1+z)$. SEAMBHs as a new
kind of cosmic ladders can be in principle extended to high-$z$ 
quasars\cite{Negrete2018,Martinez2018} from the local scaling relation\cite{Du2016a}.

Moreover, the present SARM analysis provides a direct method without the calibration issues 
of known cosmic ladders or extinction corrections. 
Future GRAVITY observations of low-$z$ SEAMBHs will provide an efficient way of calibrating 
SEAMBHs into the high-$z$ Universe. It is then expected to use AGNs for accurate measurements 
of distances from low-$z$ to high-$z$ Universe. Calibrations by different ladders are not 
necessary in the current approach.

As inquired by one of referees, we applied the simplest estimation of 3C 273 distances 
through torus image and NIR RM data. 
The angular size of its torus is $\Delta \theta_{\rm tor}=0.29\,$mas is from Keck interferometer 
observations\cite{Kishimoto2011} and the linear size $\Delta R_{\rm tor}\approx 1.0\,$lt-yr from 
the long term $K$-band light curve and the optical-UV\cite{Soldi2008}. We have a distance of 
$D_{\rm A}=\Delta R_{\rm tor}/\Delta\theta_{\rm tor}\approx 212\,$Mpc, which is smaller 
a factor of 2.6 than the present determination and the $\Lambda$CDM model, but it is still 
encouraging. We note that the $K$-band light curve has too poor cadences after 1995 to reliably 
determine $\Delta R_{\rm tor}$ (see their Figure 2 in Ref.\cite{Soldi2008}), 
moreover, this estimation needs to be improved through the sophisticated scheme to obtain 
$\Delta R_{\rm tor}$ and $\Delta\theta_{\rm tor}$ for distances like in Ref.\cite{Honig2014}. 
It should be noted that NIR continuum RM campaigns are usually much longer than the H$\beta$
ones.

\subsection{Uncertainties from the BLR model.}
Geometrical measurements of both GRAVITY and the RM data depend on the BLR model, which is assumed 
to be a geometrically thick disc with Keplerian rotation. More complicated structures and kinematics 
of the BLRs are possible, however, we should note that they can be observationally tested for
systematic errors, which can be independently estimated from observational data 
of RM campaigns through comparing with mock data.

\subsection{\it Differences compared with sole GRAVITY measurements.} 
As shown in Table 1 and Supplementary Figure\, 3, 
some results from the SARM analysis are significantly different from that from analysis of 
sole GRAVITY data, however they seem to be reasonable. As shown by Equation (\ref{eq:phase}), 
the phase curves are more sensitive to the BLR information projected
to the baseline direction. RM of broad H$\beta$ line mainly delivers its information along LOS. 
With different sensitivities of GRAVITY and RM data on model parameters,
{
the joint analysis gets benefits from that some degeneracies of parameters are 
broken by the simultaneous application of both databases} and
hence generates more robust results. For example, the SARM analysis 
yields $\bhm=5.78_{-0.88}^{+1.11}\times 10^{8}\sunm$ with relative errors smaller than that 
of $\bhm=2.6\pm 1.1\times 10^{8}\sunm$ from the pure SA analysis\cite{Sturm2018} (we noted
a $\sim 2.0\sigma$ $\bhm$-tension here). 
The SARM-measured $\bhm$ is also more reliable than the simple virial mass of black hole 
($M_{\rm vir}$)\cite{Zhang2019}. The reasons are obvious: $M_{\rm vir}$ depends on the virial 
factor, while the factor is determined by calibrating with $\bhm-\sigma$ relation\cite{Onken2004}.
Moreover, H$\beta$ lags (even its error bars are very small) generated by cross correlation 
analysis remain quite ambiguous in their exact physical meanings since the lags do not fully
cover the regions corresponding to its H$\beta$ FWHM. Currently, the lags of broad emission 
lines are explained as an emissivity-averaged radius of the BLR, however, 
the geometric radius can be specified after fixing a physical model. Once given 
the BLR geometry and 
kinematics, a joint analysis of GRAVITY and RM data generates the BLR parameters of the 
best fittings for both independent datasets (GRAVITY and RM data) simultaneously. 
For 3C 273, both $D_{\rm A}$ and $\bhm$ are quite robust [see the circle shapes of contours
in Supplementary Figure 3 
except for $(\beta,\mathscr{F})$, $(\RBLR,\mathscr{F})$
and $(\bhm,\theta_{\rm opn})$]. On the other hand, we expect SARM observations of more 
targets to better understand the differences between the SARM with others (such as the slight 
$\bhm$-tension).

\subsection{\it Geometry and kinematics.}
Roughly speaking, symmetric profiles of broad emission lines form from axisymmetric discs.
The characterized BLR model of a geometrically thick disc with Keplerian rotation can be 
observationally tested by repeating campaigns (conveniently in principle). Velocity-resolved
delays independent of models, or the 2D delay maps obtained from maximum entropy 
method\cite{Horne1991} can directly justify the BLR geometry by comparing them with the 
shapes of known geometries, such as inflows or outflows\cite{Welsh1991,Xiao2018}. 
Moreover, the Keplerian rotating disc follows a simple relation of 
$\tau_{\rm BLR}\propto V_{\rm FWHM}^{-2}$, as in NGC 5548 and others mentioned previously.
It is thus feasible to estimate the uncertainties 
of BLR geometry and kinematics contributed to systematic error bars of the model parameters.

We should note that GRAVITY measured the Pa$\alpha$ line region whereas the 10-yr RM 
campaign observed the H$\beta$. A fully self-consistent scheme should employ the same line
for GRAVITY and RM campaign, however, an RM campaign of near infrared emission lines is much 
harder than H$\beta$ line. As argued in the main text, we in principle expect that Pa$\alpha$ 
and H$\beta$ regions should be the same in the simplest model. However, $V_{\rm FWHM}$ of 
H$\beta$ and Pa$\alpha$ lines are slightly different, implying the two-line regions 
may mismatch due to different optical depths for H$\beta$ and Pa$\alpha$ photons. A simple 
estimation can be done by following.
Supposing the averaged radius ($R_{0}$) of the BLR with a difference $\Delta R$ for the two lines, 
we have $(R_{0}+\Delta R)V^{2}_{\rm Pa\alpha}=(R_{0}-\Delta R)V^{2}_{\rm H\beta}$ from 
the assumption that the two-line regions are vrialized, where $V_{\rm Pa\alpha,H\beta}$ are FWHMs of 
their profiles. We obtain the relative difference of $\Delta R/R_{0}=(q-1)/(q+1)$,
where $q= \left(V_{\rm H\beta}/V_{\rm Pa\alpha}\right)^{2}$. According to the long campaign of
3C 273, we have $V_{\rm H\beta}\approx (3100-3300)\kms$ changing with luminosity\cite{Zhang2019} while 
$V_{\rm Pa\alpha}\approx (2700-3000)\kms$ from GRAVITY\cite{Sturm2018} and NASA Infrared Telescope 
Facility (IRTF) observation\cite{Landt2008}. Taking the averaged values of 
$V_{\rm Pa\alpha,H\beta}\approx (2800,3200)\kms$, we have $\Delta R/R_{0}= 13\%$. This error bar
is comparable to that of distance measurements, but it can be greatly eliminated in the
RM$^{\rm 2D}$ modeling. 
Additionally, statistics show that 
Pa$\alpha$ line shares the same region with H$\beta$ line\cite{Landt2013}.
Ideally, RM should monitor the same emission line with GRAVITY in order to reduce the potential
differences between RM and GRAVITY detections. A NIR-RM campaign 
of monitoring Pa$\alpha$ of 3C 273 is then expected for this goal. 

On the other hand, the current analysis of the RM$^{\rm 1D}$+GRAVITY can be extended to
that of the RM$^{\rm 2D}$+GRAVITY through allowing for different but partially overlapped 
regions for the two lines, which includes information of variations of H$\beta$ 
profiles in the analysis of RM data\cite{Pancoast2014,Li2018}. 
In such a modeling scheme, the profile variations will be moderately sensitive 
to the vertical structure of the BLR. Since
Pa$\alpha$ line share the partially same regions with the H$\beta$, we relax $\mathscr{F}$ 
and $\RBLR$ to cover the two regions for different profiles of the two lines.
However, the two regions share the same geometry and kinematics, 
such as, the same $\theta_{\rm opn}$ and kinematics. Moreover, parameters describing 
cloud properties, such as anisotropy of line emissions, will be included in the 
RM$^{\rm 2D}$ modeling\cite{Pancoast2014,Li2018,Williams2018}.
Though new parameters are added in RM$^{\rm 2D}$ modeling, there are about 300 profiles
of H$\beta$ spectra of 3C 273 available to set more constraints on the parameters,  
providing opportunities to have 
better measurements of black hole mass and distance. This is the major 
contents of a separate paper of the joint analysis of GRAVITY and RM$^{\rm 2D}$ modeling 
BLR for distances.

\subsection{\it Radiation pressure.}
Motion of BLR clouds could be affected by radiation pressure\cite{Marconi2008,Netzer2010}, 
which changes kinematics of the clouds. Considering that the pressure decreases with the 
square of the distance to the black hole (for the simplest version of the pressure), 
we can combine its effects into the gravitational 
potential so that we can get the effective mass of the black hole in such a case. The effective
mass of the black hole could be slightly smaller than the present, 
but the distance to observers remains the same. 

Dimensionless accretion rates of AGNs can be estimated by
$\dot{\mathscr{M}}_{\bullet}=20.1\left(\ell_{44}/\cos i_{0}\right)^{3/2}M_{7}^{-2}$,
where $\dot{\mathscr{M}}_{\bullet}=\dot{M}_{\bullet}/\dot{M}_{\rm Edd}$, $\dot{M}_{\bullet}$
and $\dot{M}_{\rm Edd}=L_{\rm Edd}/c^{2}$ are accretion rates of the black hole and Eddington 
rate, respectively, $L_{\rm Edd}$ is the Eddington luminosity, $\ell_{44}$ 
is the 5100\AA\, luminosity in units of $10^{44}\ergs$
and $M_{7}=\bhm/10^{7}\sunm$. The $\dot{\mathscr{M}}_{\bullet}$ is derived from the
standard accretion disc model\cite{Wang2014b}. Taking $\ell_{44}=84.3$ from our RM 
campaign\cite{Zhang2019} and $M_{7}=57.8$ from the SARM analysis, we have 
$\dot{\mathscr{M}}_{\bullet}\approx 4.7$ in 3C 273, implying a slight super-Eddington 
accretor compared with others\cite{Du2018}. In such a case, the radiation pressure on cloud's 
motion is not serious for mass estimations\cite{Netzer2010}. Actually, the pressure 
can be included in the joint analysis of MCM modeling for more accurate measurements of black 
hole masses (or effective mass), fortunately, distance determinations escape from its influence. 
We would point out that the SARM analysis provides the most accurate mass of black holes in 
type I AGNs so far, and we will include radiation pressure to improve SARM analysis further
for discussions related with coevolution of SMBHs and their hosts.

\subsection{\it BLR variations.}
BLRs are known to vary with luminosities, such as in NGC 5548 (Figures 6, 8, 13 in 
Ref.\cite{Peterson1999,Lu2016,Pei2017}, respectively). 
Since GRAVITY observations only take a few hours, which are much shorter than durations  
of RM campaigns (denoted as $\Delta t_{\rm RM}$ which is usually from a few months to years), 
GRAVITY may detect the changing BLR at different epochs from the RM campaigns. This may be 
one origin of the systematic uncertainties. For 3C 273, its RM 
campaign is as long as 10yr whereas GRAVITY observations just took snaps 
over a few hours or so in the last two years. The timescale of BLR variation is roughly
given by $\Delta t_{\rm BLR}\approx R_{\rm BLR}/V_{\rm FWHM}=42\,\tau_{150}V_{3000}^{-1}$\,yr
for dynamical changes, where $\tau_{150}=\tau_{\rm BLR}/150{\rm days}$ and 
$V_{3000}=V_{\rm FWHM}/3000\kms$. Considering $\Delta t_{\rm RM}\ll \Delta t_{\rm BLR}$, 
we think that the current SARM analysis of 3C 273 avoids potential variations of the BLR 
in last ten years (the ionization fronts are fast changing with luminosity as 
$\Delta \RBLR/\RBLR\approx 0.5 \Delta L_{5100}/L_{5100}$ and it is $\lesssim 7\%$ 
from Supplementary Figure 5). 
In order to avoid systematic errors from the BLR variations, we should do RM campaigns in the same 
period of GRAVITY observations in future SARM projects. It is fortunate that GRAVITY observations 
of 3C 273 was covered by our campaign, efficiently reducing this influence on the present results.

One question may be asked if the RM-measured BLR is the same with GRAVITY measured 
regions because RM only measures the variable parts of the BLR and GRAVITY does the 
entire. This problem can be justified by comparing the mean and the RMS spectra for differences.
For 3C 273, fortunately, it is clear that the RMS shape is very similar to its mean spectrum as shown 
in Figure 2 in Ref.\cite{Zhang2019}. On the contrary, if the RMS is very distinguished  from 
the mean spectra, GRAVITY measured BLR may be very different from the RM parts making the SARM analysis
elusive. In practice, GRAVITY observations can be scheduled once at the beginning and ending epochs
to find if the BLR changes. This strategy of observations can avoid the mis-matched measurements 
between GRAVITY and RM. 

\subsection{\it Degrees of ordered motion.}
Spectroastrometry measures the mean centers of $\lambda$-photons from the BLR and 
thus depends on its angular momentum distributions, namely on the degree ($\calR$) 
of ordered motion of the BLR clouds\cite{Stern2015,Rakshit2015,Songsheng2019}. 
Fully ordered motion of BLR clouds, $\calR=1$, is presumed in Ref.\cite{Sturm2018}
about GRAVITY observations of 3C 273. In 
principle, $\calR$ should be treated as a free parameter in fitting the DPCs of
GRAVITY data, but it strongly degenerates with the BLR sizes. $\calR$ could be 
estimated from the position angles of polarized spectra arising from scattering by 
hot electrons in the mid-plane\cite{Smith2005,Songsheng2018}. Building up the relation 
between $\calR$ and position angles of the polarized spectra will be very 
helpful to understand the ${\cal R}_{0}$ parameter of BLR clouds, but we have to  
introduce more parameters describing the electron scattering zone. This is much 
beyond the scope of the present paper, but we will treat this problem separately.

\subsection{\it Error budgets.}
Error sources are generally from several aspects discussed in previous sections
in additional to the measurements of GRAVITY 
observations and RM campaigns (for ideal SARM observations, namely, focus on the same line). 
First, the degree of disordered motion could contribute 
uncertainties to the present analysis. This involves the formation of the BLR, either from 
disc winds\cite{Czerny2011} or tidally captured clumps from a dusty torus\cite{Wang2017}. 
We expect to test $\calR$ through polarized spectra. Second, non-Keplerian 
kinematics and a non-disc BLR could be other sources of systematic errors. Fortunately, 
this can be evaluated by repeating RM campaigns of individual AGNs to demonstrate the 
sources through direct test of the relation between $\tau_{\rm H\beta}-{\rm FWHM}$ 
and velocity-resolved delay relation ($\tau_{\rm H\beta}-V$) for the BLR geometry,
where $V$ is the velocity bin. Actually, the systematic errors can be efficiently 
alleviated through selecting targets with symmetric profiles of broad emission lines.
Third, the quasi-simultaneous observations of GRAVITY and RM campaigns might
imply different regions measured by the two tools. However, this could be avoided
if the SARM observations can be performed under reasonable schedule of observations
or within a BLR dynamical timescale. Fourth, selection of radial and angular distributions 
of BLR clouds in the model is another source of systematic errors for both GRAVITY and RM 
data. In order to quantitatively issue the errors, we need to simulate models to show
error bars. Detailed discussions on this problem are beyond the 
scope of this paper, but we leave it in a future work.

Finally, for individual 3C 273, some errors are from observations of the RM campaign.
There are a few key valleys and peaks for the 
determinations of model parameters, however, they are in season gaps, such as the 
second half years of 2012, 2015, 2016, 2017. Accuracies of model parameters are 
then affected in the joint analysis. Actually, we continue the campaign with a goal 
of getting more diagnostic valleys and peaks in next few years.

\subsection{Future SARM projects.} 
Spectroastrometry and reverberation mapping must come together for excellent 
studies of AGN physics and cosmology, but also for close-binaries of supermassive 
black holes (CB-SMBHs)\cite{Wang2018,Songsheng2019} radiating nano-Hertz gravitational 
waves (to be detected by the Pulsar Timing Arrays\cite{Burke2019}). 
Only GRAVITY observations of 3C 273 have been reported so far\cite{Sturm2018}, 
but similar targets selected from existing catalogs are listed for future SARM project
in the Supplementary Table 1. GRAVITY will install new grisms in October 2019 so that 
$K\approx 11.0-11.5$ targets can be observed as routine research of AGNs. We 
note most the targets with redshifts $z\lesssim 0.08$, in which only Br$\gamma$ 
line (is significantly fainter than Pa$\alpha$ line) could be detected by GRAVITY.
{Simulations for 3C 120 ($K=10.78$, see Supplementary Table 1) 
showed very promising detections of the Br$\gamma$ line, which was given by a talk 
of M. R. Stock. It can be found from website of the workshop
({\tt https://www.torus2018.org/TALKS/18.12.10-S3.3-Stock.pdf}). 
From page 15 of the talk's file, we find that Br$\gamma$ interferometric signals 
are strong enough for a joint analysis in the future. On the other hand, 
Br$\gamma$ line forms
from a transition of electrons from $n=7$ to 4. In such a context, we should understand 
differences of Br$\gamma$ and H$\beta$ line regions through CLOUDY and  test 
results from the RM$^{\rm 2D}$+GRAVITY modeling.}
 we thus expect that the 
current GRAVITY observes the targets listed here in near future. Optical spectra 
of all the targets are given in Supplementary Figure 6. 
The targets actually cover three kinds of AGNs: 1) \feii-strong objects; 2) candidates 
of CB-SMBHs appearing with H$\beta$ asymmetry; 3) \feii-weak objects. 

A brief strategy of the future SARM projects could be as outlined below.
RM campaigns of the targets are expected to perform: 1) reveal velocity-resolved delays 
for kinematics and geometry (with physical sizes); 2) construct 2-dimensional 
transfer functions for justifying CB-SMBH candidates; 3) show stability of the 
BLR from multiple campaigns, for the goal of establishing BLR properties including
stability of their structures. 
This needs much work of 2m telescopes to prepare for GRAVITY observations. After then,
GRAVITY observations provide differential phase curves for structure, kinematics and 
angular sizes of the BLR. On the other hand, RM of GRAVITY near-infrared emission 
lines will, in principle, provide fully self-consistent data for the joint analysis.
In such a campaign, targets should be selected more carefully to avoid NIR absorptions 
of the atmosphere. The campaigns could be conducted through 4m-class telescopes.
We outline three aspects below (but they are valid for RM campaigns of H$\beta$ line 
with different broad emission lines in NIR). 

SARM observations are for scientific goals as followings.
Targets will be spatially resolved by GRAVITY observations along with RM campaigns 
to explore signatures of flattened rotating disc, inflows, or outflows in BLR in 
order to study accretion process and formation of the BLR connecting with dusty 
torus\cite{Wang2017}. In particular, 
those of optical \feii-strong AGNs (about 1/3 of PG quasars\cite{Boroson1992}) 
usually have smaller BLRs compared with objects with the same luminosities (significantly 
downward deviating from the well-known $R-L$ relation)\cite{Du2018}, and most of them are 
super-Eddington accreting massive black holes (SEAMBHs). SARM-based measurements of
SEAMBHs might reveal more details of BLR structure and kinematics as well as 
physics of super-Eddington accretion process. Moreover, 
super-Eddington accretion process as a key phase across cosmic time is a critical step 
toward fast growth of seed black holes to form SMBHs in high-$z$ 
Universe\cite{Volonteri2005,Wang2006,Milo2009,Regan2019}. This is very compelling for
the increasing large samples of DESI (Dark Energy Survey Instrument) to tackle growth 
and formation issues of the SMBHs. Third, SEAMBHs are suggested to be a new kind of cosmic 
candles for the high-$z$ cosmology\cite{Wang2013,Wang2014b,Marziani2014}. SARM-based 
measurements of SEAMBHs will build up precision ladders approaching to high-$z$ Universe.

With the unprecedented power of high-spatial resolutions, GRAVITY offers opportunity of
spatially resolve CB-SMBHs. AGNs with asymmetric profiles have likely more 
complicated BLR structures, and some of them could 
be CB-SMBH candidates\cite{Wang2018,Nguyen2019}. 
A dedicated project of Monitoring AGNs with H$\beta$ Asymmetry (MAHA) is being conducted to 
construct 2-dimensional transfer functions\cite{Wang2018,Songsheng2019a,Kovacevic2019}  
through the Wyoming Infrared Observatory 2.3m telescope\cite{MAHA1} for the 
CB-SMBHs of future GRAVITY observations. The MAHA project
is expected to provide the most promising candidates of the CB-SMBHs for GRAVITY, and for 
a joint analysis of the SARM data\cite{Songsheng2019}. We would like to point out that
Ark 120, Mrk 704 and several others show interesting features of CB-SMBHs in their 2D transfer 
functions. They are good targets of GRAVITY observations with obvious signatures of differential
phase curves\cite{Songsheng2019b} in near future. Moreover, it could
be possible to measure orbital parameters (such as masses of component black holes, inclination
and ellipticity of the orbits) by the joint analysis of SARM data
in order to predict properties of low-frequency gravitational waves radiated by the CB-SMBHs.  

Optical \feii-weak AGNs are usually sub-Eddington accretors\cite{Du2018}. We select some of them
with stable and flattened-Keplerian disc-like BLR guaranteeing physical conditions of the SARM 
targets to efficiently reduce systematic errors of distance measurements of AGNs. They are 
excellent targets of GRAVITY observations for cosmology outlined in this paper. Fortunately 
again, this can be done by RM campaigns using 2m telescopes. From Supplementary Table 1, 
we have about $N\approx (30, 50)$ targets with $K\lesssim (11.0,11.5)$, respectively. This makes 
it feasible to establish a future SARM project for the $H_{0}$-measurements, providing 
$\Delta H_{0}/H_{0}\lesssim (3,2)$ per cent, respectively, which are valid for a test of the 
current $H_{0}$-tension. With one 4yr SRAM project for $\sim 100$ AGNs, we can achieve $\sim 1.5\%$.
If the uncertainties 
($\sim\!13\%$ in 3C 273) 
are included, $N\approx 200$ could be necessary. However, in principle, 
this systematic errors can be alleviated by RM$^{\rm 2D}$+GRAVITY measurements, hence, 
the $N\approx 200$ should be significantly reduced. On the other hand, with capability of future 
interferometers in space (spatial resolution up to $\sim 1\mu$as by the {\it SIM}-Lite 
for optical $\sim 20$mag AGNs\cite{Unwin2008,Ding2009}), we can accurately measure all information 
of the H$\beta$ line region (angular and linear sizes) in optical bands for low-$z$ quasars so that 
this systematic errors completely vanish. 
The $H_{0}$-measurements can be done with unprecedented precisions by an AGN sample much smaller 
than the present estimates. 
GRAVITY$^{+}$ with much improved sensitivity\cite{Widmann2018} is expected 
to explore high-$z$ cosmology (i.e., the Hubble parameter) through the present SARM scheme.
In this case, one needs to use larger telescope (e.g., 5m class) to monitor GRAVITY+ targets 
in order to avoid too long RM campaigns. It should be stressed again that this
is geometric for high-$z$ cosmology.

In a brief summary, the near future SEAMBH-, CB-SMBH-, and $H_{0}$-SARM projects will significantly 
advance understanding of AGN physics, close-binaries of supermassive black holes for nano-Hertz 
gravitational waves, and precision cosmology, respectively. It is highly desired to perform 
the three feasible projects in a few years.

{\bf Data Availability.} The data that support the plots within this paper and
other findings of this study are available from the corresponding author upon reasonable request.

{\bf Code Availability.} All the codes used in this paper are available from the corresponding
author upon reasonable request.

\vglue 0.5cm

\end{methods}

\clearpage

\begin{center}{\large\bf Supplementary Information}\end{center}
\vglue -0.5cm
In this Section, we provide all figures used in Methods, target selection and target list
including name, coordinates and spectra for future SARM observations.

\subsection{The Figures} are
\begin{itemize}
\item Carton of the broad-line regions to illustrate the current model used in this paper
(Supplementary Figure 1).
\item Fittings of the differential phase curves observed by GRAVITY (Supplementary Figure 2).
\item Results of probability distributions of physical parameters in the fittings (Supplementary Figure 3).
\item Light curves of $\gamma$-rays and $V$-band (Supplementary Figure 4).
\item Light curves of optical continuum and H$\beta$ line since 2009 (Supplementary Figure 5).
\end{itemize}

\subsection{SARM target selection}
%

We selected targets from existing catalogs of AGNs. They 
are from the 2dF\cite{si2dF}, 6dF\cite{si6dF} (also available from 
{\tt http://vizier.u-strasbg.fr/viz-bin/VizieR}), 
the Veron Catalog of Quasars 
and Active Galactic Nuclei\cite{siVeron2010}, the Hamburg/ESO Survey\cite{siWisotzki2000}, 
and the Quasar Catalog of Sloan Digital Sky Surveys\cite{siParis2018}. Two criteria 
are used for selection: 1) coordinates ${\rm Dec.}\le 20^{\circ}$ for GRAVITY of 
the VLTI (we assume that 2m telescopes are available for the SARM project in both 
southern and northern hemispheres); 2) $K$-band magnitudes $K<11.5$ for the GRAVITY 
(private communications with E. Sturm). Supplementary Table \ref{tab:targets} 
lists the targets with necessary 
information including H$\beta$ lags if measured, or estimated from the normal $R-L$ 
relation\cite{siBentz2013}, angular sizes of the BLR according to the standard 
$\Lambda$CDM model ($H_{0}=67\,{\rm km\,s^{-1}\,Mpc^{-1}}$, $\Omega_{\Lambda}=0.315$ 
and $\Omega_{\rm M}=0.685$). All targets listed here are type I AGNs. Future 
observations of spectroscopic surveys of 4MOST ({\tt https://www.4most.eu/cms/}) 
will greatly increase numbers of targets for the SARM project. Supplementary
Table \ref{tab:targets} 
is useful both for GRAVITY observations and RM campaigns for RM community over the 
world to make the SARM analysis in future. We also provide one single epoch of all
the targets in Supplementary Figure 6, but it should be noted that profiles 
of the H$\beta$ line change generally.

\clearpage

\begin{figure*}
    {\centering
    \includegraphics[width= 1.\textwidth,trim=-20 170 -10 100]{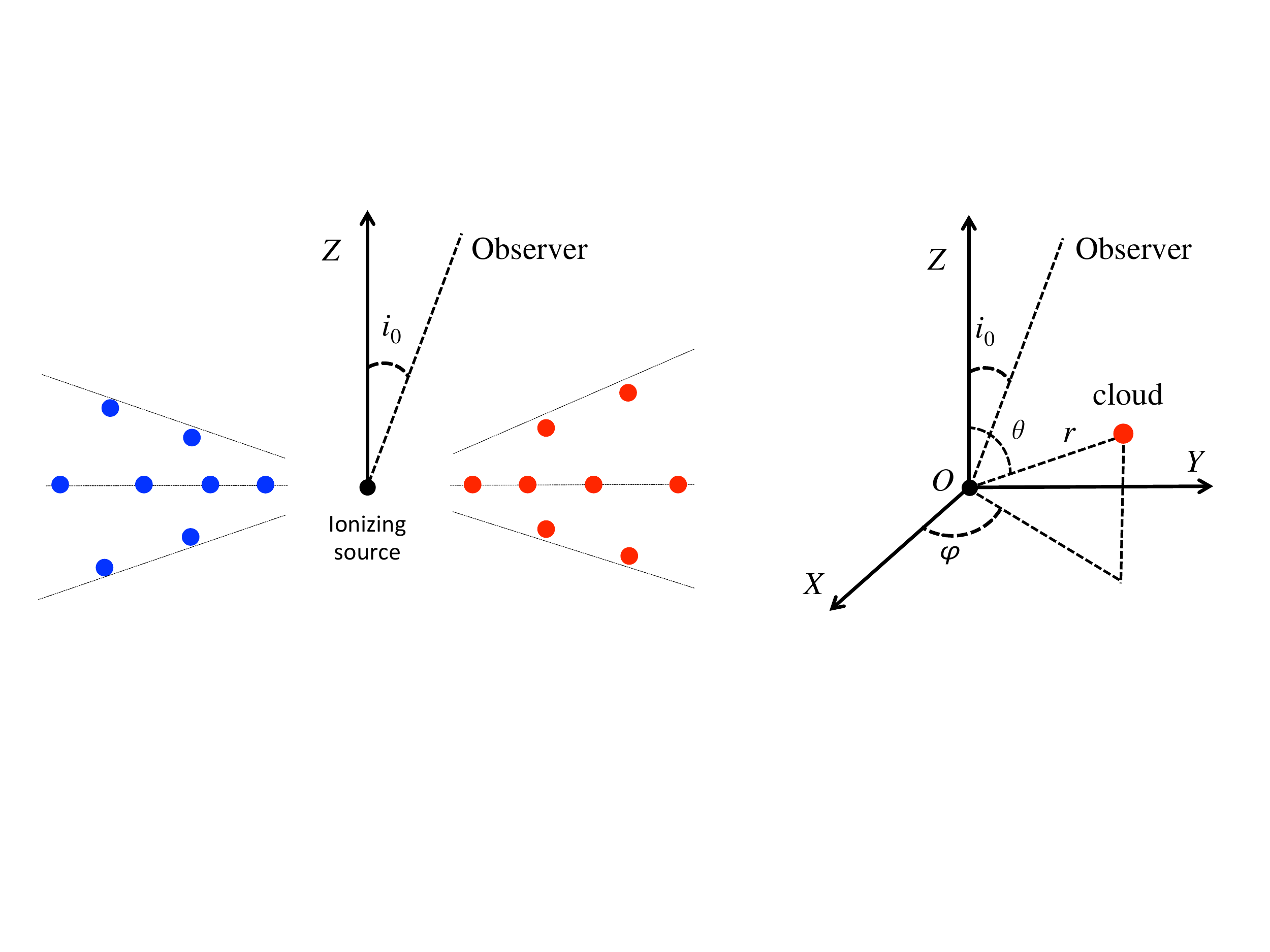}
    {\footnotesize {\bf Supplementary Figure 1. Structure, geometry and kinematics of the 
    simplest BLR in AGNs.} It is
    characterized by a flattened disc with opening angle of $\theta_{\rm opn}$. 
    The left panel is a cartoon illustration of the BLR.
    Clouds are presumed to be optically thin and are orbiting around the central black hole 
    with Keplerian velocity. The blue and red 
    clouds are approaching and receding to observers, respectively. A remote observer has an 
    inclination of $i_{0}$ located in the $O-YZ$ plane. The right panel is the coordinate system 
    we used. Here $r$ is the distance of the cloud to the central SMBH, 
    while $\theta$ and $\phi$ are polar and azimuthal angle of the cloud respectively.}}
\label{fig:BLR}
\end{figure*}

\begin{figure*}
    {\centering
    \includegraphics[scale = 0.67]{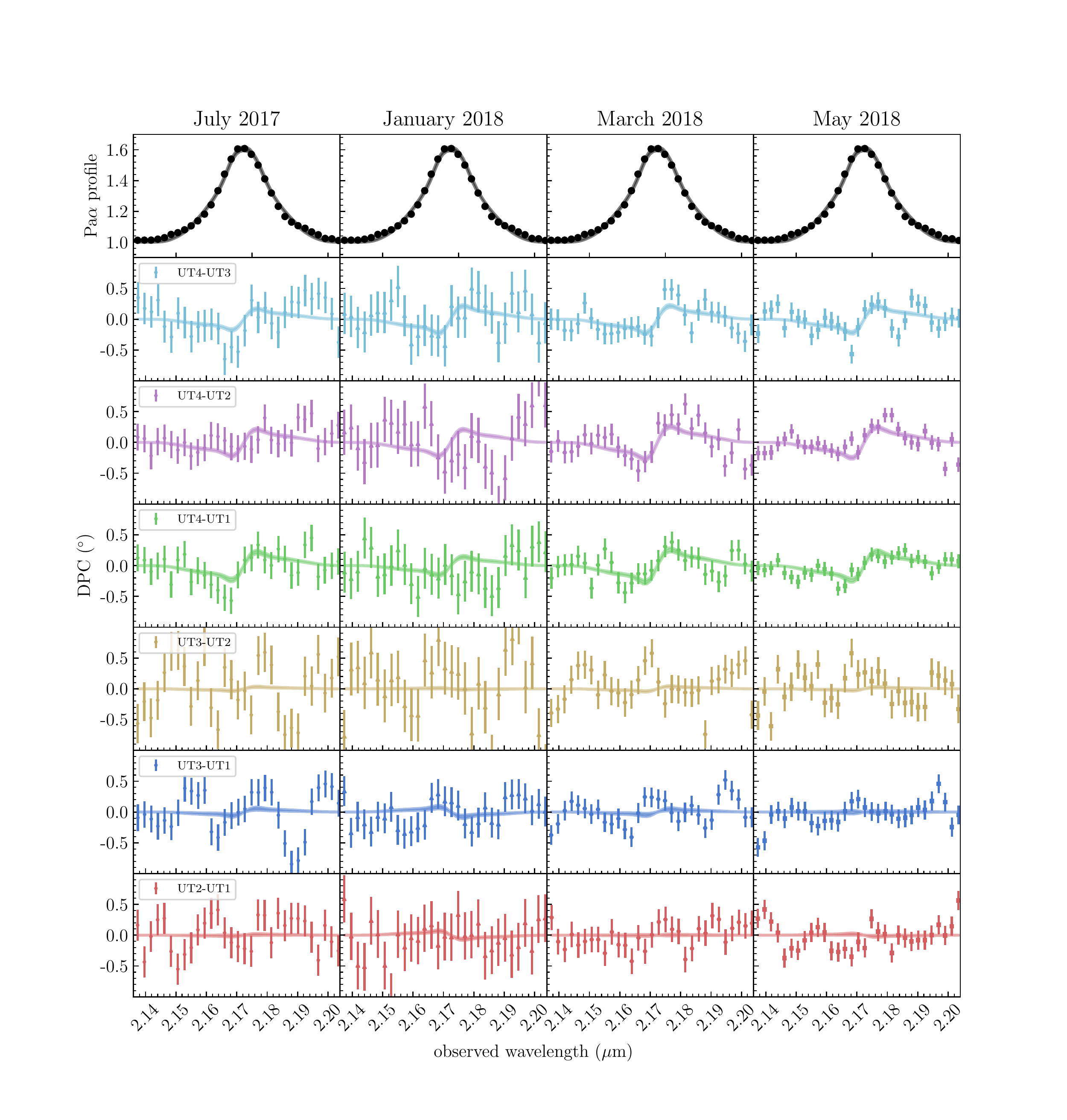}
    \vglue -1cm
    {\footnotesize {\bf Supplementary Figure 2.
    Fittings of the DPCs from the 24 baseline interferometric data} (the reduced $\chi^{2}=1.33$).
    Different colors indicates DPCs from different telescope pairs. 
    The first rows are Pa$\alpha$ line profiles taken from averaged profiles of
    the four epochs observed. The DPCs generate the angular sizes of the BLR.
    Error bars for all data points reflect $1\sigma$ uncertainties.}}
\label{fig:phase-curves}
\end{figure*}
\begin{figure*}
    {\centering
    \includegraphics[scale = 0.325]{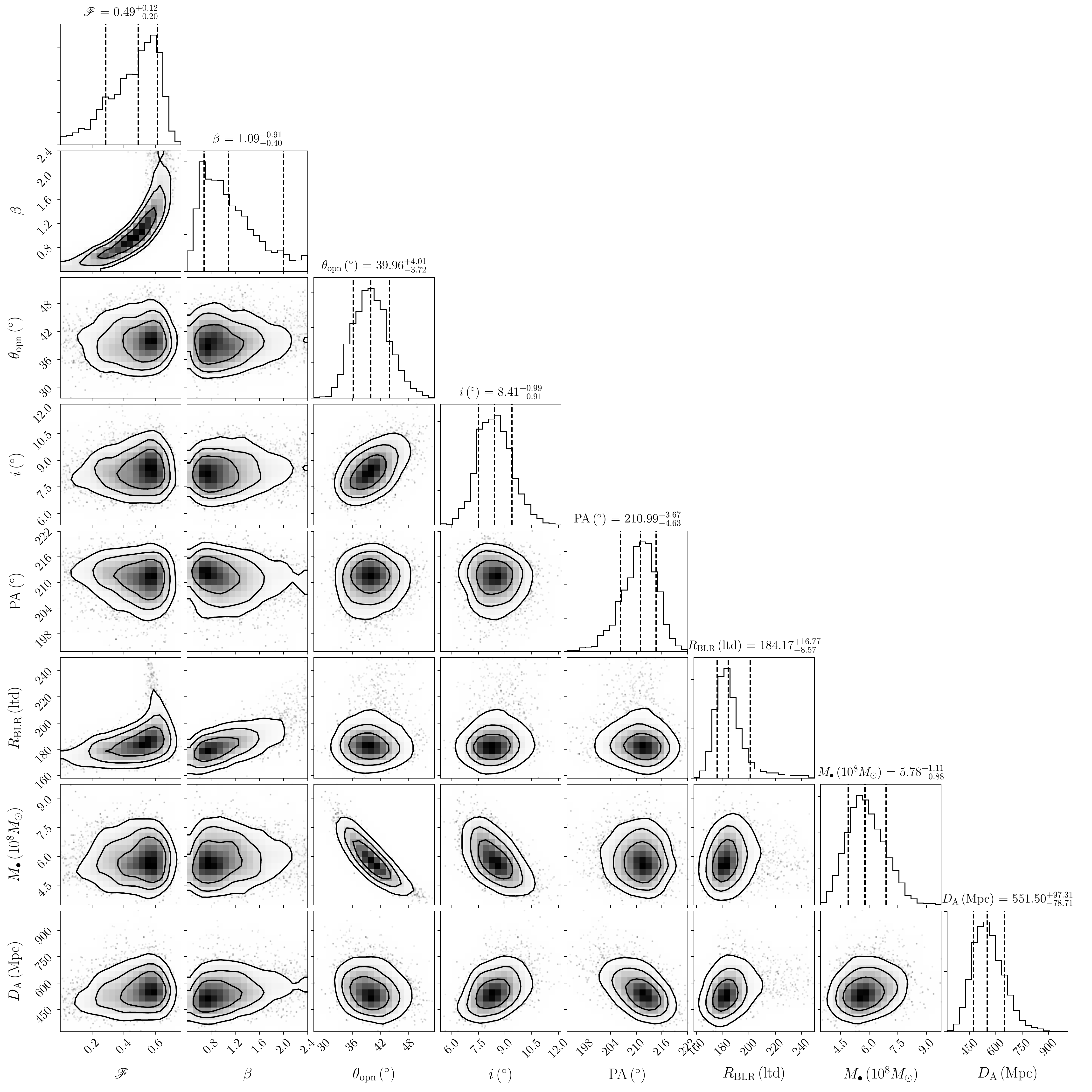}
    {\footnotesize {\bf Supplementary Figure 3.
    Corner plot of the BLR parameters and angular distances.} Probability density 
    distributions and contours of parameters are from the joint fittings of 
    differential phase curves and H$\beta$ reverberation data. The best values of 
    parameters are given on the tops of panels. Error bars are quoted 
    at the $1\sigma$ level, which are given by each parameter's distributions.
    The dashed lines in the one-dimensional distributions are the 
	$16\%$, $50\%$ and $84\%$ quantiles, and contours are at $1\sigma$, $1.5\sigma$ 
	and $2\sigma$, respectively.}}
\label{fig:contour-results}
\end{figure*}

\clearpage

\begin{figure*}
    {\centering
    \includegraphics[scale = 0.6]{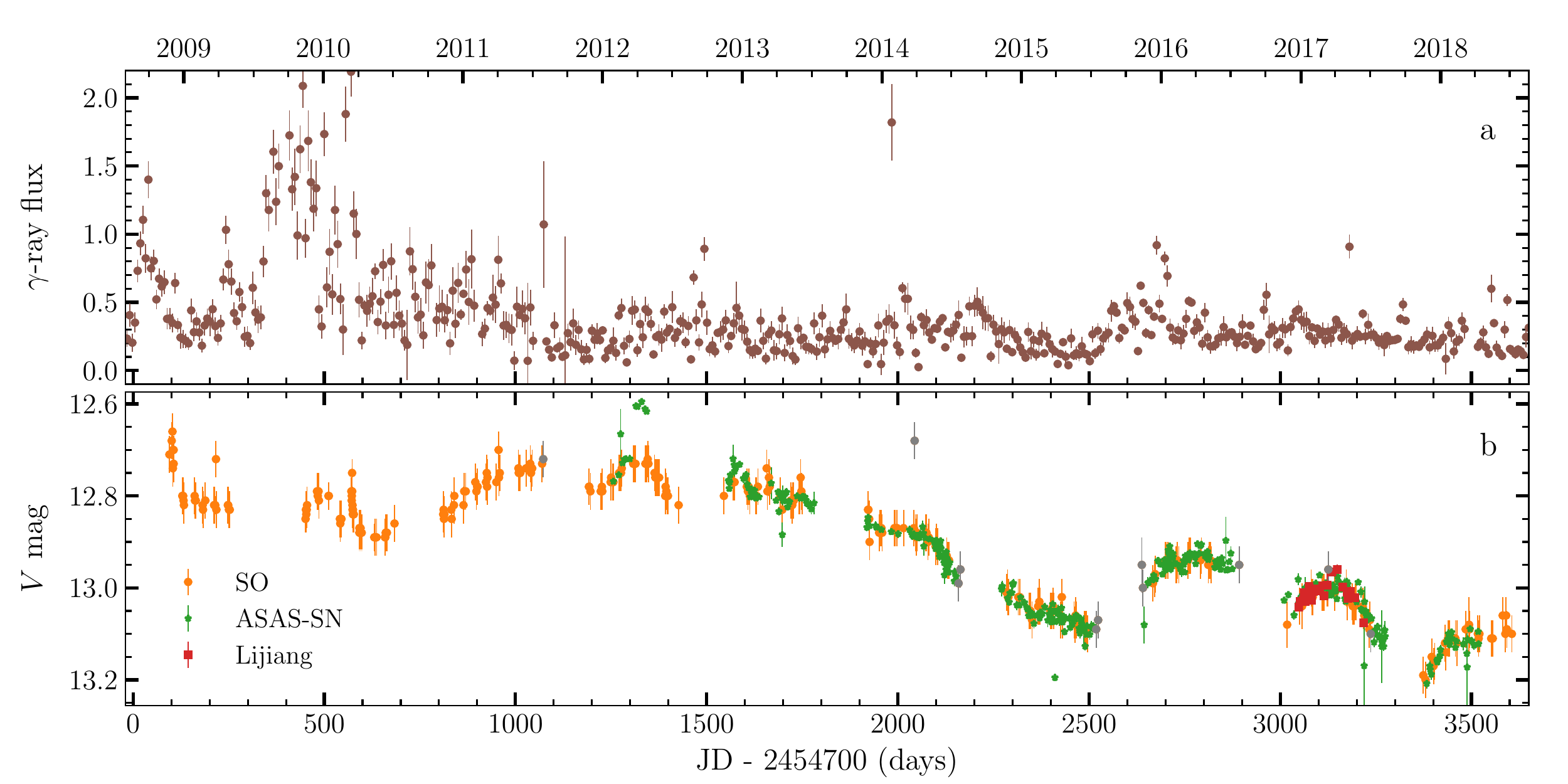}
    {\footnotesize {\bf Supplementary Figure 4.
    The {\it Fermi} $\gamma$-ray light curve and comparison with $V$-band observed for 10 yrs.}
    ASAS-SN: All-Sky Automated Survey for 
    Supernovae (see http://www.astronomy.ohio-state.edu/~assassin/index.shtml).
    Error bars for all data points reflect $1\sigma$ uncertainties. 
	The different colors of data points in panel b indicates different sources of V 
	magnitude data, as shown in the lower left corner of the panel.}}
\label{fig:gamma-ray}
\end{figure*}

\begin{figure*}
    {\centering
    \includegraphics[scale = 0.8]{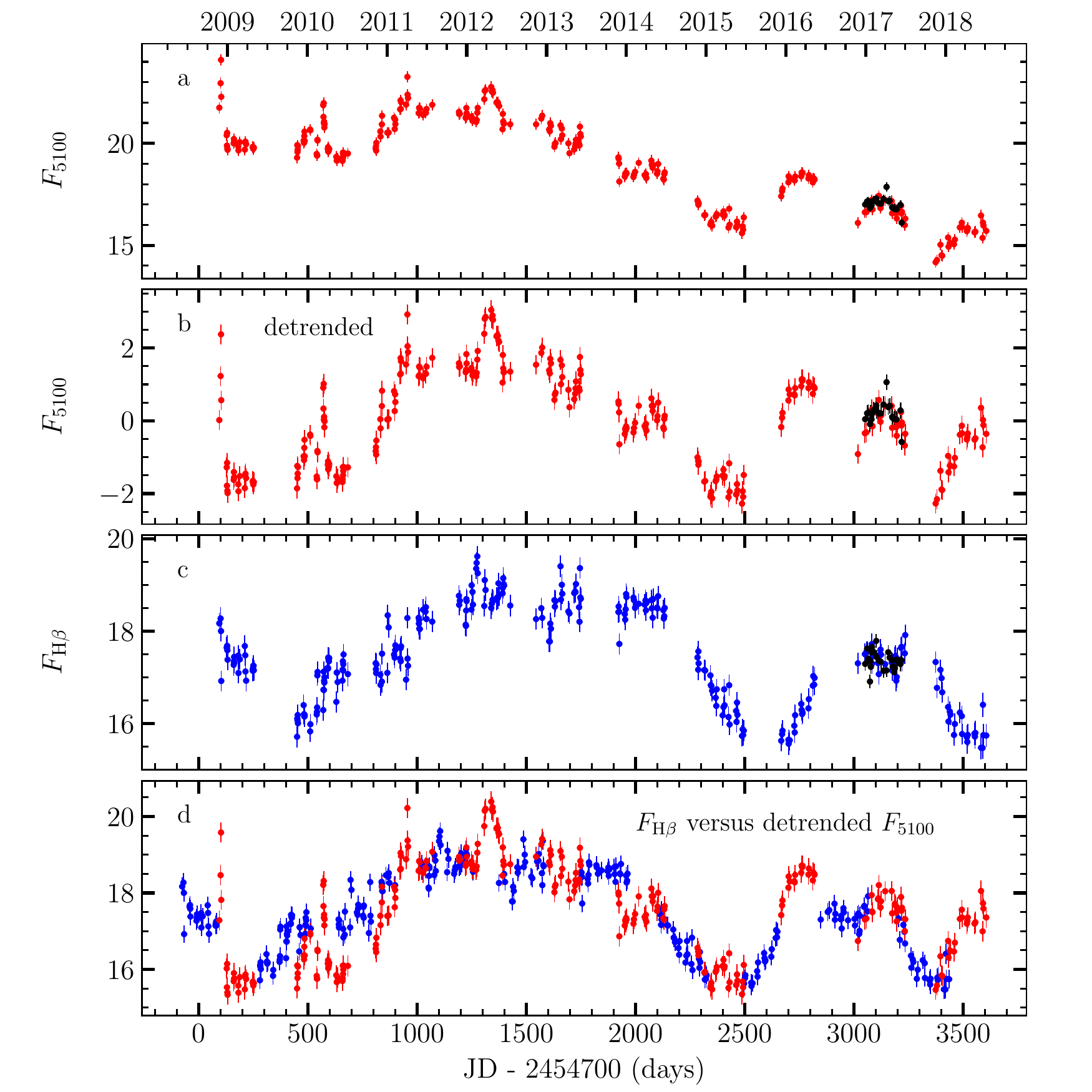}
    {\footnotesize {\bf Supplementary Figure 5.
    The light curves of continuum and H$\beta$ since 2009.} Panel {\it a}: 5100\AA\,
    (in units of ${\rm 10^{-15}\, erg\, s^{-1}\, cm^{-2}\AA^{-1}}$),
    {\it b}: de-trended 5100\AA, {\it c}: H$\beta$, and {\it d} H$\beta$ versus 
    de-trended 5100\AA. In panel {\it d}, the H$\beta$ shifted light curve
    is multiplied by a scaling factor for a simple comparison. The H$\beta$ line
    between 2009-2012 has poor response to the varying continuum. Error bars for all 
    data points reflect $1\sigma$ uncertainties. Red data points are light curves of 
    continuum, while blues ones are those of H$\beta$. The black points in panels 
    ({\it a})-({\it c}) are obtained from the Lijiang 2.4m telescope after
    inter-calibrations.
    }}
\label{fig:tenyr}
\end{figure*}

\clearpage

\renewcommand{\thefootnote}{\fnsymbol{footnote}}
\renewcommand{\arraystretch}{0.81}

\begin{center}{\normalsize Supplementary Table 1 \\ Targets of Future SARM Observations}\end{center}
{\scriptsize
\begin{longtable}{llclccccccc}\hline\hline
No. & Name & R.A.& Declinations &$z$   & $K$  & $V$  &$\log L_{5100}$&$\RBLR$&$\xiBLR$& Ref.\\ \hline
    &     &     &              &       &     &      & ($\ergs$)       & (ltd)  & ($\mu$as) &      \\ \hline
1.  & PG 0003+199/Mrk 335$^\ddagger$  & 00h06m19.52s & +20d12m10.5s & 0.026 & 10.54 & 14.33 & 43.72 & 10.6 - 16.8   & 16.6 - 26.3 & $R_{1}$  \\ \relax
2.  & PG 0007+106/III Zw 2$^\dagger$  & 00h10m31.01s & +10d58m29.5s & 0.089 & 11.72 & 15.47 & 44.37 & 12.6          & 6.1         & $R_{2}$  \\ \relax
3.  & PG 0050+124/I Zw 1$^\ddagger$   & 00h53m34.94s & +12d41m36.2s & 0.059 & 10.31 & 14.04 & 44.57 & 37.2          & 26.5        & $R_{3}$  \\ \relax
4.  & Fairall 9                       & 01h23m45.78s & -58d48m20.8s & 0.047 & 10.95 & 13.77 & 44.48 & 17.4          & 15.3        & $R_{4}$  \\ \relax
5.  & Mrk 1018                        & 02h06m15.99s & -00d17m29.2s & 0.042 & 11.33 & 14.31 & 44.17 & 41.3          & 40.1        &          \\ \relax
6.  & Mrk 590/NGC 863$^*$             & 02h14m33.56s & -00d46m00.1s & 0.026 & 10.40 & 15.68 & 43.20 & 14.0 - 29.2   & 21.4 - 44.7 & $R_{5}$  \\ \relax
7.  & Mrk 1044$^\ddagger$             & 02h30m05.52s & -08d59m53.3s & 0.016 & 10.81 & 13.94 & 43.49 & 4.8           & 11.6        & $R_{6}$  \\ \relax
8.  & Mrk 1048/NGC 985$^\dagger$      & 02h34m37.77s & -08d47m15.4s & 0.043 & 10.84 & 14.02 & 44.30 & 48.6          & 46.4        &          \\ \relax
9.  & HE 0343-3943                    & 03h45m12.53s & -39d34m29.3s & 0.043 & 11.13 & 14.64 & 44.04 & 35.5          & 34.2        &          \\ \relax
10. & 2MASS J04145265-0755396         & 04h14m52.67s & -07d55m39.9s & 0.038 & 10.89 & 14.75 & 43.90 & 29.7          & 31.9        &          \\ \relax
11. & 3C 120$^\dagger$                & 04h33m11.10s & +05d21m15.6s & 0.033 & 10.51 & 14.15 & 44.01 & 20.2 - 38.1   & 24.9 - 47.0 & $R_{7}$  \\ \relax
12. & Mrk 618                         & 04h36m22.24s & -10d22m33.8s & 0.036 & 10.90 & 14.10 & 44.10 & 37.9          & 43.5        &          \\ \relax
13. & Ark 120/Mrk 1095$^\dagger$      & 05h16m11.42s & -00d08m59.4s & 0.033 & 10.14 & 14.59 & 43.83 & 16.2 - 70.0   & 20.2 - 87.1 & $R_{8}$  \\ \relax
14. & MCG -02-14-009                  & 05h16m21.18s & -10d33m41.4s & 0.028 & 11.13 & 15.50 & 43.34 & 15.0          & 21.3        &          \\ \relax
15. & 2MASX J05580206-3820043         & 05h58m02.00s & -38d20m04.7s & 0.034 & 10.00 & 14.80 & 43.77 & 25.5          & 30.7        &          \\ \relax
16. & 2MASS J06235520+0018433         & 06h23m55.18s & +00d18m42.9s & 0.094 & 11.10 & 15.75 & 44.31 & 49.1          & 22.9        &          \\ \relax
17. & NGC 2617                        & 08h35m38.79s & -04d05m17.6s & 0.014 & 11.65 & 14.00 & 43.33 & 14.9          & 41.6        &          \\ \relax
18. & Mrk 704$^\dagger$               & 09h18m26.01s & +16d18m19.2s & 0.029 & 10.68 & 14.13 & 43.91 & 30.3          & 42.0        & $R_{9}$  \\ \relax
19. & PG 0923+129/Mrk 705             & 09h26m03.29s & +12d44m03.6s & 0.029 & 11.16 & 14.24 & 43.87 & 28.6          & 39.7        &          \\ \relax
20. & Mrk 1239                        & 09h52m19.10s & -01d36m43.5s & 0.020 & 9.69  & 14.27 & 43.52 & 18.7          & 37.6        &          \\ \relax
21. & NGC 3227$^\dagger$              & 10h23m30.58s & +19d51m54.2s & 0.004 & 9.73  & 12.48 & 42.81 & 3.8           & 38.2        & $R_{10}$ \\ \relax
22. & HE 1029-1401$^\dagger$          & 10h31m54.30s & -14d16m51.0s & 0.086 & 11.14 & 14.08 & 44.89 & 100.6         & 50.8        &          \\ \relax
23. & ESO 265- G 023                  & 11h20m48.01s & -43d15m50.4s & 0.057 & 11.49 & 14.65 & 44.29 & 47.8          & 35.4        &          \\ \relax
24. & PG 1126-041/Mrk 1298$^\ddagger$ & 11h29m16.66s & -04d24m07.6s & 0.062 & 11.11 & 14.33 & 44.50 & 61.9          & 42.1        &          \\ \relax
25. & NGC 3783                        & 11h39m01.76s & -37d44m19.2s & 0.010 & 9.83  & 12.60 & 43.56 & 10.2          & 41.5        & $R_{11}$ \\ \relax
26. & PG 1211+143$^\ddagger$          & 12h14m17.67s & +14d03m13.1s & 0.081 & 11.29 & 14.62 & 44.62 & 73.3          & 39.1        & $R_{12}$ \\ \relax
27. & PG 1226+023/3C 273$^\ddagger$   & 12h29m06.70s & +02d03m08.6s & 0.158 & 9.99  & 12.90 & 45.94 & 146.8 - 306.8 & 43.7 - 91.2 & $R_{13}$ \\ \relax
28. & Mrk 1330/NGC 4593               & 12h39m39.43s & -05d20m39.3s & 0.009 & 9.82  & 12.60 & 43.50 & 3.7 - 4.3     & 16.4 - 18.9 & $R_{14}$ \\ \relax
29. & ESO 323- G 077                  & 13h06m26.13s & -40d24m52.8s & 0.015 & 9.32  & 13.22 & 43.69 & 23.1          & 61.3        &          \\ \relax
30. & 2MASX J13411287-1438407         & 13h41m12.90s & -14d38m40.6s & 0.042 & 11.25 & 14.36 & 44.13 & 39.7          & 39.1        &          \\ \relax
31. & IC 4329A                        & 13h49m19.27s & -30d18m34.0s & 0.016 & 9.25  & 13.54 & 43.62 & 15.6          & 38.8        & $R_{15}$ \\ \relax
32. & ESO 511- G 030                  & 14h19m22.42s & -26d38m41.0s & 0.022 & 10.96 & 13.93 & 43.76 & 25.0          & 44.9        &          \\ \relax
33. & PG 1426+015/Mrk 1383            & 14h29m06.59s & +01d17m06.5s & 0.087 & 11.07 & 14.25 & 44.83 & 93.6          & 46.9        &          \\ \relax
34. & PG 1501+106/Mrk 841$^\dagger$   & 15h04m01.20s & +10d26m16.2s & 0.036 & 11.39 & 14.24 & 44.06 & 36.3          & 40.8        &          \\ \relax
35. & 2MASX J15115979-2119015         & 15h11m59.80s & -21d19m01.7s & 0.045 & 10.92 & 14.76 & 44.03 & 35.0          & 32.4        &          \\ \relax
36. & 2MASS J16461038-1124042         & 16h46m10.39s & -11d24m04.2s & 0.074 & 11.07 & 16.20 & 43.91 & 30.2          & 17.4        &          \\ \relax
37. & 2MASS J17050039-0132286         & 17h05m00.39s & -01d32m28.6s & 0.030 & 11.09 & 15.12 & 43.55 & 19.4          & 25.9        &          \\ \relax
38. & PDS 456                         & 17h28m19.80s & -14d15m55.9s & 0.184 & 9.83  & 14.33 & 45.52 & 216.2         & 56.9        &          \\ \relax
39. & Fairall 51                      & 18h44m53.98s & -62d21m53.4s & 0.014 & 10.20 & 13.94 & 43.36 & 15.3          & 42.8        &          \\ \relax
40. & ESO 141- G 055                  & 19h21m14.14s & -58d40m13.1s & 0.037 & 10.64 & 13.70 & 44.29 & 48.2          & 53.2        & $R_{16}$ \\ \relax
41. & 2MASX J19373299-0613046         & 19h37m33.01s & -06d13m04.8s & 0.010 & 10.53 & 13.53 & 43.24 & 13.2          & 50.9        &          \\ \relax
42. & NGC 6814                        & 19h42m40.64s & -10d19m24.6s & 0.005 & 9.81  & 12.97 & 42.87 & 6.6           & 50.1        & $R_{17}$ \\ \relax
43. & 2MASX J19490928-1034253         & 19h49m09.28s & -10d34m25.0s & 0.024 & 10.22 & 13.93 & 43.82 & 27.0          & 45.3        &          \\ \relax
44. & NGC 6860                        & 20h08m46.89s & -61d06m00.7s & 0.015 & 10.31 & 13.26 & 43.67 & 22.5          & 60.1        &          \\ \relax
45. & 2MASS J20304171-7532430         & 20h30m41.63s & -75d32m42.8s & 0.114 & 11.43 & 15.00 & 44.79 & 88.6          & 34.7        &          \\ \relax
46. & MC 2031-307                     & 20h34m31.35s & -30d37m28.8s & 0.019 & 10.90 & 13.30 & 43.88 & 29.2          & 60.3        &          \\ \relax
47. & Mrk 509$^\ddagger$              & 20h44m09.74s & -10d43m24.5s & 0.034 & 10.19 & 13.54 & 44.29 & 79.6          & 94.4        & $R_{18}$ \\ \relax
48. & 2MASX J21090996-0940147         & 21h09m09.97s & -09d40m14.7s & 0.027 & 10.87 & 13.85 & 43.94 & 31.2          & 47.6        &          \\ \relax
49. & [HB89] 2121-179                 & 21h24m41.64s & -17d44m46.0s & 0.112 & 11.38 & 15.20 & 44.69 & 78.5          & 31.4        &          \\ \relax
50. & PG 2130+099/Mrk 1513$^\ddagger$ & 21h32m27.81s & +10d08m19.5s & 0.063 & 10.63 & 14.46 & 44.46 & 22.9          & 15.3        & $R_{19}$ \\ \relax
51. & PG 2214+139/Mrk 304             & 22h17m12.26s & +14d14m20.9s & 0.066 & 11.28 & 14.41 & 44.52 & 63.5          & 40.9        &          \\ \relax
52. & MR 2251-178                     & 22h54m05.80s & -17d34m55.0s & 0.064 & 11.22 & 14.02 & 44.65 & 74.6          & 49.3        &          \\ \relax
53. & NGC 7469/Mrk 1514$^\dagger$     & 23h03m15.62s & +08d52m26.4s & 0.016 & 9.63  & 12.84 & 43.92 & 4.5 - 10.8    & 11.0 - 26.4 & $R_{20}$ \\ \hline
\label{tab:targets}
\end{longtable}
\vglue -1.5cm

$^{*}$NGC 863 has another name of Mrk 590, which is known as a changing-look AGN, and recently it 
begins to activate again. It would be interesting to measure the BLR history from type I to type II.
\vglue -0.5cm

Targets marked with $\dagger$ are being conducted by the MAHA project, and with $\ddagger$ are 
being conducted by the SEAMBH project of the Lijiang and the Carlo Alto (CAHA) telescopes.
\vglue -0.5cm

$V$-band magnitudes of most targets are given by ASAS-SN Photometry Database\cite{siKochanek2017}, except:
PG 0003+199 (Ref.\cite{siHaas2011}); 
Mrk1018 (Ref.\cite{sideVaucouleurs1988}); 
Mrk590, Ark120 (Ref.\cite{siBentz2009b});
MCG -02-14-009, NGC 2617, MC 2031-307 (Ref.\cite{siVeron2010}).
$K$-band magnitudes of all targets are taken from 2MASS measurements\cite{siSkrutskie2006}.
Most of them are measured using $4''$ aperture, except Mrk 1048 and NGC 6814 using 
$14.0''\times14.0''$ aperture.
\vglue -0.5cm

$L_{5100}$ (in units of ${\rm ergs\,s^{-1}}$) is estimated using $V$-band magnitude and luminosity distances, 
assuming a composite quasar spectra from the SDSS\cite{siVandenBerk2001} for all objects.
\vglue -0.5cm

For those mapped targets, $\RBLR$ is from measured time lags or the minimum and maximum lags (if there
are multiple campaigns). Mrk 704 and  ESO 141- G 055 were mapped, but failed to measure reliable lags 
due to the poor quality of the data. For targets without mapping campaigns, lags are estimated by the 
$R-L$ relation\cite{Bentz2013} for AGN.
Angular size for all target are calculated through $\xi_{\rm BLR}=R_{\rm BLR}/D_{\rm A}$.
\vglue -0.5cm

References of mapped AGNs:
$R_{1}$: Ref.\cite{siPeterson1998,siPeterson2004, siGrier2012, siDu2014};\,\,
$R_{2}$: Ref.\cite{siGrier2012};\,\,
$R_{3}$: Ref.\cite{siHuang2019};\,\,
$R_{4}$: Ref.\cite{siWinge1996,siSantos-Lleo1997};\,\,
$R_{5}$: Ref.\cite{siPeterson1998,siPeterson2004};\,\,
$R_{6}$: Ref.\cite{siWang2014b};\,\,
$R_{7}$: Ref.\cite{siPeterson1998,siPeterson2004,siGrier2012,siKollatschny2014,siMAHA1};\,\,
$R_{8}$: Ref.\cite{siWinge1996,siPeterson1998,siPeterson2004,siDoroshenko2008,siMAHA1};\,\,
$R_{9}$: Ref.\cite{siPeterson1998};\,\,
$R_{10}$: Ref.\cite{siDenney2010};\,\,
$R_{11}$: Ref.\cite{siStirpe1994};\,\,
$R_{12}$: Ref.\cite{siKaspi2000,siPeterson2004};\,\,
$R_{13}$: Ref.\cite{siKaspi2000, siZhang2019};\,\,
$R_{14}$: Ref.\cite{siDenney2006,siBarth2013};\,\,
$R_{15}$: Ref.\cite{siWinge1996};\,\,
$R_{16}$: Ref.\cite{siWinge1996};\,\,
$R_{17}$: Ref.\cite{siBentz2009};\,\,
$R_{18}$: Ref.\cite{siPeterson1998,siPeterson2004};\,\,
$R_{19}$: Ref.\cite{siGrier2012};\,\,
$R_{20}$: Ref.\cite{siCollier1998,siPeterson2004,siPeterson2014};
}

\normalsize
\renewcommand{\thefootnote}{\arabic{footnote}}
\renewcommand{\arraystretch}{1.0}

\clearpage

\begin{figure*}
{    \centering
    \includegraphics[scale = 0.25]{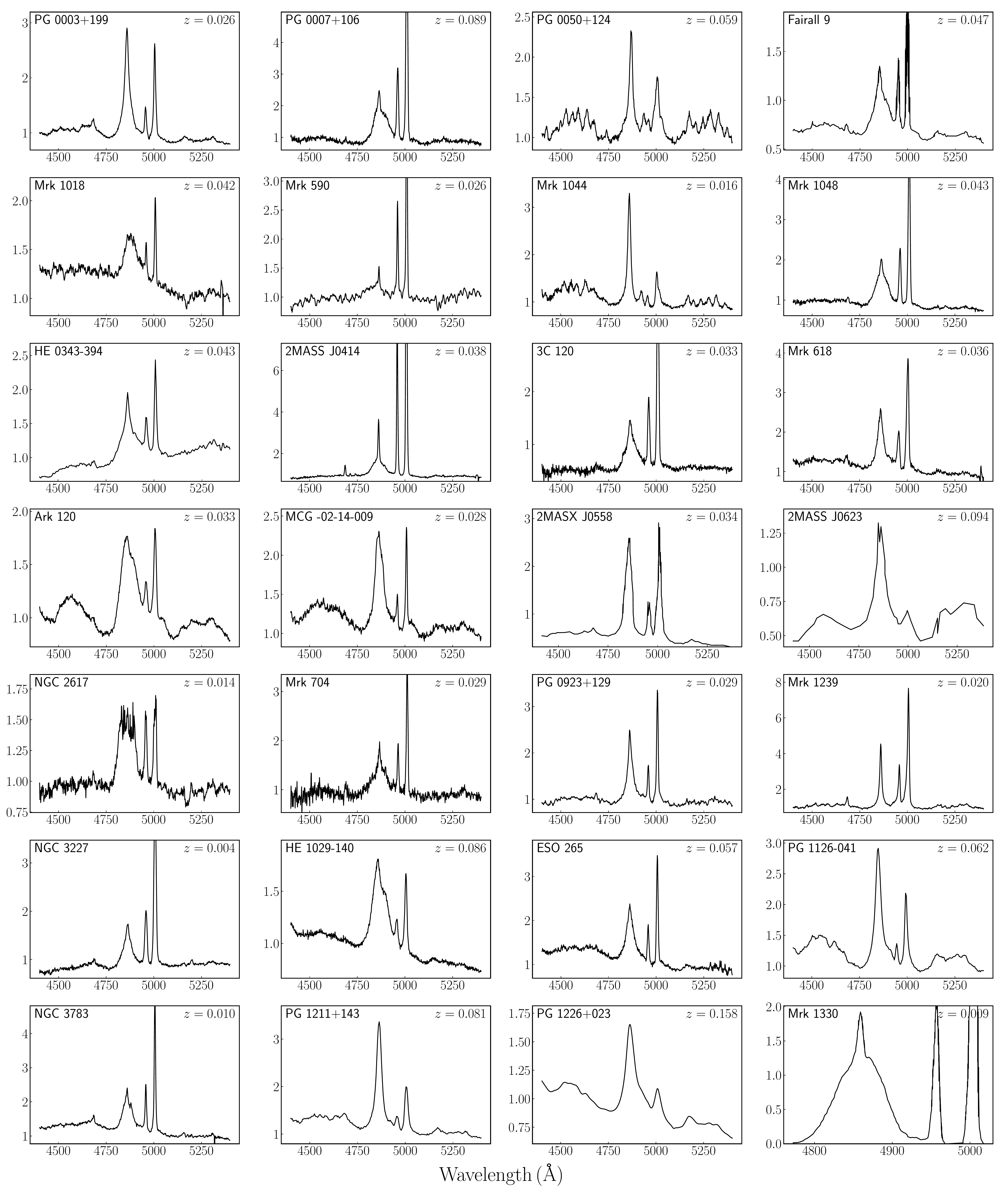}
    \vglue -0.5cm
    {\footnotesize {\bf Supplementary Figure 6.
    Spectra of all selected targets of the SARM projects in rest frame.} All the spectra 
    are from publications (see references). Spectral fluxes are in units of 
    ${\rm ergs\,s^{-1}\,\AA^{-1}}$ but scaled by arbitrary factors for convenience.  
    Ref. Of spectra: object No. 1 is from  Ref.\cite{siDu2014};
No. 2, 8, 10, 12, 17, 18 and 30 from the MAHA project;}
    }
\label{fig:spectra}
\end{figure*}

\begin{figure*}
    \centering
    \includegraphics[scale = 0.25]{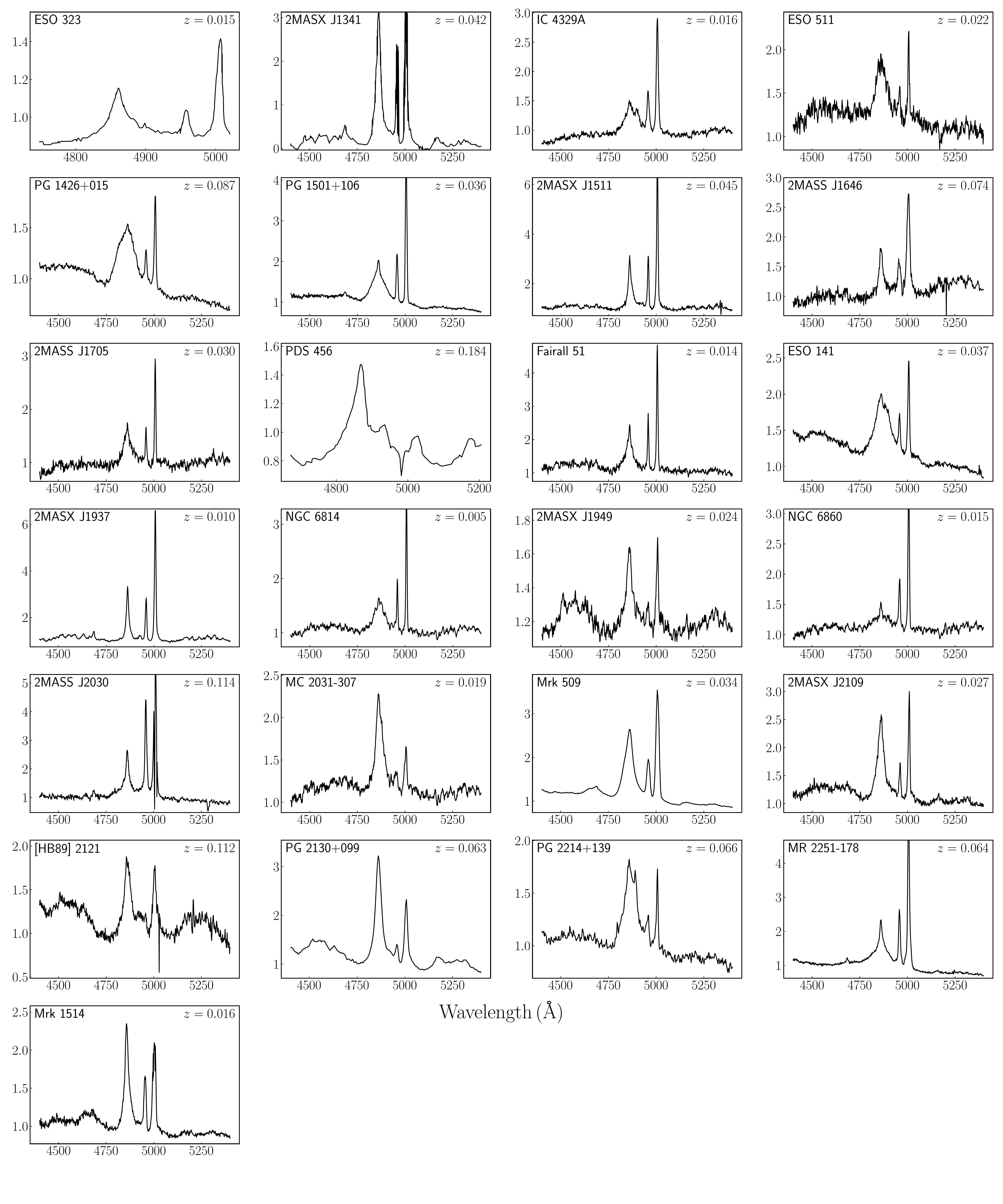}
\vglue -0.75cm
{\footnotesize   {\bf Supplementary Figure 6  {\it continued. }}
No. 3  from  Ref.\cite{siSantos-Lleo1997};
No. 6 from Ref.\cite{siWang2014b};
No. 4, 5, 7, 9, 11, 13, 16, 19, 21, 27, 28, 31-33, 35-42, 44, 45 and 48 from Ref.\cite{siJones2009};
No. 14 and 26 from Ref.\cite{siMarziani2003};
No. 15, 20, 29, 46, 47 and 49 from Ref.\cite{siBoroson1992};
No. 22 and 43 from the CAHA project;
No. 23  from  Ref.\cite{siZhang2019};
No. 24  from  Ref.\cite{siWilliams2018};
No. 25  from  Ref.\cite{siThomas2017};
No. 34  from  Ref.\cite{siSimpson1999}.}
\end{figure*}

\end{document}